\documentclass[prper,aps,twocolumn,groupedaddress,floats,showpacs,final,superscriptaddress]{revtex4-1}
\usepackage{leading}
\usepackage[latin3]{inputenc}
\usepackage[makeroom]{cancel}
\usepackage{graphicx}
\usepackage{amsmath}
\usepackage{amsfonts}
\usepackage{amssymb}
\usepackage{color}
\usepackage[left=2cm,right=2cm,top=2cm,bottom=2cm]{geometry}
\usepackage{graphicx} 
\usepackage{dcolumn}
\usepackage{bm}
\usepackage{simplewick}
\usepackage{array}
\usepackage{appendix}

\RequirePackage[
   hyperindex,colorlinks,bookmarksnumbered,
   plainpages=true,pdfstartview=FitH]{hyperref}
\hypersetup{linkcolor=blue,urlcolor=blue,citecolor=blue}
\usepackage{hyperref}
\usepackage{mathrsfs}
\definecolor{purple}{rgb}{0.5,0,0.6}

\usepackage{ulem}
\renewcommand{\emph}[1]{\textit{#1}}
\definecolor{darkblue}{rgb}{0,0,0.5}
\definecolor{darkgreen}{rgb}{0,0.5,0}
\definecolor{darkred}{rgb}{.7,0,0}
\definecolor{purple}{rgb}{0.5,0,0.6}
\definecolor{orange}{rgb}{1,0.5,0}
\definecolor{grey}{rgb}{.6,.6,.6}
\definecolor{lightpink}{rgb}{1,0.7,0.75}
\definecolor{pink}{rgb}{1,0.4,0.58}
\definecolor{deeppink}{rgb}{1,0.08,0.58}

\newcommand{\DK}[1]{{\color{black}{#1}}} 

\renewcommand{\emph}[1]{\textit{#1}}

\begin{document}

\title{Nonlinear Seebeck effect of SU($N$) Kondo impurity}



\author{D. B. Karki}
\affiliation{International School for Advanced Studies (SISSA), Via Bonomea 265, 34136 Trieste, Italy}
\affiliation{The  Abdus  Salam  International  Centre  for  Theoretical  Physics  (ICTP)
Strada  Costiera 11, I-34151  Trieste,  Italy}
\author{Mikhail N. Kiselev}
\affiliation{The  Abdus  Salam  International  Centre  for  Theoretical  Physics  (ICTP)
Strada  Costiera 11, I-34151  Trieste,  Italy}



\begin{abstract}
We develop a theoretical framework to study the influences of coupling asymmetry on the thermoelectrics of a strongly coupled SU($N$) Kondo impurity based on a local Fermi liquid theory. Applying non-equilibrium Keldysh formalism, we investigate charge current driven by the voltage bias and temperature gradient in the strong coupling regime of an asymmetrically coupled SU($N$) quantum impurity. The thermoelectric characterizations are made via non-linear Seebeck effects. We demonstrate that the beyond particle-hole (PH) symmetric SU($N$) Kondo variants are highly desirable with respect to the corresponding PH symmetric setups in order to have significantly improved thermoelectric performance. The greatly enhanced Seebeck coefficients by tailoring the coupling asymmetry of beyond PH symmetric SU($N$) Kondo effects are explored. Apart from presenting the analytical expressions of asymmetry dependent transport coefficients for general SU($N$) Kondo effects, we make a close connection of our findings with the experimentally studied SU(2) and SU(4) Kondo effects in quantum dot nano structures. \DK{Seebeck effects associated with the theoretically proposed SU(3) Kondo effects are discussed in detail. }
\end{abstract}

\maketitle




\section{Introduction}
\vspace*{-3mm}
The increasing demand of quantum technologies for energy harvesting has attracted growing attention towards the necessity for the nano material based energy converters~\cite{ld0}. The presence of quantization effects in nano scale systems allows the controllable comprehension and subsequent control of underlying transport processes~\cite{ld00}. In addition, nano scale systems can offer greatly enhanced thermoelectric response with respect to conventional bulk counterparts~\cite{ld00, ld0, ld1}. These properties of nano scale systems has rekindled the field of thermoelectricity~\cite{casti}. Over the past years, several experiments has resulted in exciting thermoelectric measurement for nano scale systems, such as quantum dots (QDs), carbon nano-tubes (CNTs), quantum point contacts (QPCs), etc~\cite{casti, Blanter, Z}. The rapid progress of nano technology has allowed the fine tuning of nano scale transport process, nonetheless, complete understanding of electron interactions in such small scale remains the most challenging task~\cite{costi1}. 

\DK{A generic nano device consists of a quantum impurity with intrinsic spin S which is tunnel coupled to two electron reservoirs, the source and the drain.} The low energy transport processes are then controlled by the strong interaction between localized spin S and itinerant electrons in the reservoirs. The spin S{=}1/2 impurity interacting with a single orbital channel of conduction electrons forms a fully screened ground state resulting in \DK{quasiparticle} resonances at the Fermi level. This paradigmatic screening phenomenon is termed as Kondo effect~\cite{kondo} which is characterized by a low energy scale $T_K$, the Kondo temperature. The many-body Kondo resonance at the Fermi level opens an effective path towards the enhancement of thermoelectric production at the nano scale level~\cite{kiselev}. Recent experiments~\cite{if3, if1, if2, if4, heiner, paulo} have further expanded the scope of transport measurements in Kondo correlated nano scale systems. Most of these studies have been focused on the transport measurement for the spin S{=}1/2 Kondo impurity described by the SU(2) symmetry group. However, the conventional SU(2) Kondo effects, being protected by particle-hole (PH) symmetry, offer vanishingly small thermoelectric conversion~\cite{costi1}. To achieve appreciable
thermopower, the occupation factor of the quantum impurity
should be integer, while the
PH symmetry should be lifted~\cite{dee1}. The SU($N$)  Kondo model with integer occupancy $m$ offers the possibility of avoiding half-filled regime so as to achieve the enhanced thermoelectric production over the conventional SU(2) Kondo correlated systems~\cite{azema,dee1, dee4}. 

The orbital degeneracy of the quantum impurity combines with the true spin symmetry to form the Kondo effect described by higher symmetry group SU($N$). Here the occupancy factor $m$ takes all possible values starting from 1 to $N{-}1$. The paradigmatic SU(4) Kondo physics has been experimentally studied in CNTs~\cite{jh, sasa0, sasa, if2, ll, ll1}, double QDs~\cite{keller} and single-atom transistor~\cite{st}. Various theoretical works~\cite{hur,su4_21, su4_22, su4_23, Lim, su4_11} have contributed towards better understanding of SU(4) Kondo physics over past years. In addition, exciting proposals \DK{have} been put forth for the experimental realization of different variants of SU($N$) Kondo systems. Possible realization of SU(3) Kondo effects using triple QDs with three and four edge states of the quantum Hall effects was suggested in Ref.~\cite{aash}, which been verified recently using numerical renormalization group study~\cite{rok1}. The proposals for the solid-state realization of SU(6)~\cite{kita} and SU(12)~\cite{su12} Kondo effects have likewise attracted considerable attentions both theoretically and experimentally. Beside obtaining the solid-state realization of these exotic SU($N$) Kondo effects,
an increasing effort has been put in their cold atomic realization~\cite{nashida1,salamon, nashida2, su6}. 

Most of the previous studies on SU($N$) Kondo effects have been focused solely on charge current measurements. However, thermoelectric characterization in a generic nano device usually involves the Seebeck effects. To the best of our knowledge very few studies have tried to uncover the thermoelectric measurements of Kondo effects described by higher symmetry group. The Seebeck effects with a SU(4) Kondo effects have been studied in Ref.~\cite{azema} and a general theoretical framework for thermoelectric transport of a SU($N$) Kondo model has been developed recently in Ref.~\cite{dee1}. These studies are limited to the setup with perfectly symmetrical tunnel coupling, which is very rarely the case of an experiment. In fact, junction asymmetry could provide important informations about the underlying many body effects~\cite{mora1}, thus, has to be taken into account in any
calculation to compare its result with the experimental data~\cite{ll3}. Therefore unveiling the effects associated with the coupling asymmetry towards the thermoelectric characterization, the Seebeck effects, of a SU($N$) Kondo effects has remained a challenging problem for many years. In this contribution we develop a theoretical framework based on a local Fermi-liquid theory in combination with the out of equilibrium Keldysh approach to study the influences of coupling asymmetry on the thermoelectric transport of a strongly coupled SU($N$) Kondo impurity. 

The paper is organized as follows. In Sec.~\ref{model1} we discuss in detail on the formulation of the model, for thermoelectric transport calculations at the strong-coupling regime of SU($N$) Kondo effects, capturing the effects of coupling asymmetry and arbitrary temperature and chemical potentials of the electron reservoirs. We outline the charge current calculations for a SU($N$) Kondo impurity which account for both
elastic and inelastic effects using the nonequilibrium Keldysh
formalism in Sec.~\ref{current}. The Sec.~\ref{results} is devoted to the summary of our results for the thermoelectric transport coefficients of SU($N$) Kondo correlated systems characterizing the non-linear Seebeck effects. In this section, apart from presenting the analytical expressions of coupling asymmetry
dependent transport coefficients for general SU(N ) Kondo effects, we made a separate analysis of thermoelectrics with i) experimentally studied SU(2) and SU(4) Kondo effects and 
ii) theoretically proposed SU(3) Kondo effects. The last section~\ref{conclusion} contains the conclusion of our work together with the possible future \DK{research plans} based on the present work.
\begin{figure}[t]
\includegraphics[scale=1.6]{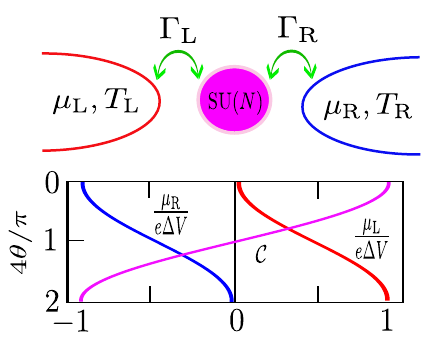}
\caption{Upper panel: Schematic representation of experimental setup for investigating Seebeck effect in nanostructures, where a SU($N$) quantum impurity is sandwiched between two conducting reservoirs. The left (red) and right (blue) reservoirs are in thermal equilibrium, separately, at temperature $T_{\rm L}$ and $T_{\rm R}$ respectively. The tunneling-matrix elements from the impurity to the left/right reservoirs are characterized by $t_{\rm L}{=}t\cos\theta$ and $t_{\rm R}{=}t\sin\theta$ with $\theta\in(0, \pi/2)$. Lower panel: The asymmetry of the tunneling junction is accounted for by introducing a parameter $\mathcal{C}\equiv(\Gamma_{\rm L}-\Gamma_{\rm R})$$/$$(\Gamma_{\rm L}+\Gamma_{\rm R}){=}\cos2\theta$ with $\Gamma_{\rm L/R}{=}\pi \rho_{\rm res} |t_{\rm L/R}|^2$, $\rho_{\rm res}$ being the density of states of the reservoirs. The magenta line represents the variation of asymmetry parameter $\mathcal{C}$ with respect to the asymmetry angle $\theta$. We choose the Fermi level in such a way that
the chemical potentials of the left and right reservoirs take some specific values $\mu_{\rm {L/R}}{=}\pm \frac{e\Delta V}{2}(1\mp\mathcal{C})$. This choice of chemical potentials amounts to greatly simplify the calculation of charge and heat current through a strongly coupled Kondo impurity (see text for detail). These chemical potentials are represented by the red and blue curve respectively.}\label{setup1}
\vspace*{-5mm}
\end{figure}
\vspace*{-5mm}
\section{Model description}\label{model1}
\vspace*{-3mm}
We consider a quantum impurity tunnel coupled to two conducting reservoirs as shown in Fig.~\ref{setup1}. The impurity possess $N$-fold degeneracy by combining the spin and other degrees of freedom, such as the orbital degeneracy. In addition, there are $N$-species (orbitals) of electrons in both left (L) and right (R) reservoirs. The rotation of the reservoir's electrons is then described by the SU($N$) transformation. Therefore, to describe our system we start form SU($N$) impurity Anderson model~\cite{and1, and2},
\begin{equation}\label{sana1}
H{=}\sum_{k,\rm r}\varepsilon_k\left[ c^{\dagger}_{{\rm L}, k\rm r}c_{{\rm L}, k\rm r}+c^{\dagger}_{{\rm R}, k\rm r}c_{{\rm R}, k\rm r}\right]+H_{\rm imp}+H_{\rm tun}.
\end{equation}
Here we introduce the notation ``${\rm r}$" to represent the orbital index that takes all possible values starting from 1 to $N$. The operator $c^{\dagger}_{{\rm \gamma}, k\rm r}$ creates an electron with
momentum $k$ in $\rm r$-th orbital of the $\gamma\;({=}{\rm L, R})$ reservoir. The energy of conduction electrons $\varepsilon_k$ is measured with
respect to the chemical potential $\mu$. The second term of Eq.~\eqref{sana1} represents the
Hamiltonian of the impurity possessing $N$ degenerate flavors with single energy level $\varepsilon_d$. Then we write the impurity Hamiltonian as
\begin{equation}\label{sane3}
H_{\rm imp}=\varepsilon_d\sum_{\rm r}d^{\dagger}_{\rm r}d_{\rm r}+U\sum_{\rm {r<r'}}
d^{\dagger}_{\rm r}d_{\rm r} d^{\dagger}_{\rm r'}d_{\rm r'},
\end{equation}
where $d^{\dagger}_{\rm r}$ is the electron creation operator of the impurity and $U$ represents the charging energy which is assumed to be the largest energy scale of the model. The tunneling processes from the impurity to the reservoirs are accounted for by the very last term of the Eq.~\eqref{sana1},
\begin{equation}\label{sane4}
H_{\rm tun}=\sum_{k,\rm r} \left(t_{\rm L} c^{\dagger}_{{\rm L}, k\rm r}+t_{\rm R} c^{\dagger}_{{\rm R}, k{\rm r}}\right)d_{\rm r}+{\rm H.c.},
\end{equation}
We explicitly assume the tunneling asymmetry by assigning the tunneling-matrix elements $t_{\gamma}$ such that $t_{\rm L}{=}t\cos\theta$ and $t_{\rm R}{=}t\sin\theta$ with $\theta{\in} \left(0, \pi/2\right)$. Then the intrinsic total local level width associated with the tunneling is given by $\Gamma_{\rm \gamma}{=}\pi \rho_{\rm res} |t_{\rm \gamma}|^2$ with $\rho_{\rm res}$ being the density of states of the reservoirs. For the sake of clarity, we introduce the parameter $\mathcal{C}\equiv(\Gamma_{\rm L}-\Gamma_{\rm R})$$/$$(\Gamma_{\rm L}+\Gamma_{\rm R}){=}\cos2\theta$ to characterize the asymmetry of the tunneling junction. This asymmetry further appears in the Glazman-Raikh rotation~\cite{GR} of Eq.~\eqref{sana1} in the basis of reservoirs electrons 
\begin{equation}\label{sane5}
 \left(%
\begin{array}{c}
  b_{k \rm r} \\
  a_{k \rm r} \\
\end{array}%
\right)= \left(%
\begin{array}{cc}
  \cos\theta & \phantom{-}\sin\theta \\
  \sin\theta & -\cos\theta \\
\end{array}%
\right)\left(%
\begin{array}{c}
  c_{{\rm L}, k \rm r} \\
  c_{{\rm R}, k \rm r} \\
\end{array}%
\right).
\end{equation}
Note that the transformation Eq.~\eqref{sane5} effectively decouples the operators $a_{k\rm r}$ from the impurity degrees of freedom. Here we consider the general case of having arbitrary number of electrons $m{=}1, 2,\cdots,N-1$ in the impurity. Therefore, the specific choice of impurity level $\varepsilon_d{=}U(1{-}m{-}m/N)$ provides the fundamental representation with $\sum_{\rm r}d^{\dagger}_{\rm r}d_{\rm r}\equiv\sum_{\rm r}n_{\rm r}{=}m$. We then perform the Schrieffer-Wolff transformation~\cite{sw} followed by the rotation Eq.~\eqref{sane5} of the Hamiltonian Eq.~\eqref{sana1} to project out the charge states, which results in
\begin{equation}\label{sane6}
\mathscr{H}=\sum_{k,\rm r}\varepsilon_k\left(a^{\dagger}_{k\rm r}a_{k\rm r}+b^{\dagger}_{k\rm r}
b_{k\rm r}\right)+H_{\rm Kondo}.
\end{equation}
The Kondo Hamiltonian is expressed in terms of anti-ferromagnetic coupling $J_K$ between the impurity spin $\vec{\mathcal{S}}$ and the spin operator of reservoir electrons placed at the origin $\vec{T}$ as~\cite{t,mora2, mora1}
\begin{equation}\label{sane7}
H_{\rm Kondo}=J_K \vec{\mathcal{S}}\cdot\vec{T},\;\;\; J_K=\frac{t^2}{U}\frac{N^2}{m(N-m)}.
\end{equation}
The $N^2{-}1$ traceless components of impurity spin $\mathcal{S}^i (i{=}1, 2\cdots N^2{-}1)$ are given by $\mathcal{S}^{{\rm r, r'}}{=}d^{\dagger}_{\rm r}d_{\rm r'}{-}m/N\delta_{\rm {r, r'}}$ with the constraint ${\rm r,r'}{\neq}{N, N'}$. Likewise, the spin operator of reservoir electrons placed at the origin is expressed as $\vec{T}{=}\sum_{kk',\rm rr'}b^{\dagger}_{k\rm r}\Lambda^i_{\rm rr'}b_{k'\rm r'}$, ${\bf{\Lambda}}^i$ being the $N\times N$ generators of SU($N$) group. Note that $\mathcal{S}^i$ are $\frac{N!}{m! (N-m)!}\times\frac{N!}{m! (N-m)!}$ matrices acting on states with $m$ electrons.

The ground state of spin $S{=}1/2$ SU($N$) impurity considered in this work is characterized by the complete screening of impurity spin which results in the formation of Kondo singlet. The low-energy regime of fully-screened Kondo effect is consistently describe by FL theory~\cite{Nozieres, aff, Cox}. Applying the standard point-splitting procedure~\cite{aff, mora2, mora1, dee1} to the Hamiltonian Eq.~\eqref{sane7} imparts the low energy FL Hamiltonian of SU($N$) Kondo impurity,
\begin{eqnarray}
\mathscr{H}_{0}&=&\nu\sum_{\rm r}\int_{\varepsilon}  \varepsilon \left[ a^{\dagger}_{\varepsilon \rm r} a_{\varepsilon \rm r}+b^{\dagger}_{\varepsilon \rm r} b_{\varepsilon \rm r}\right]\label{HamFL},\\
\mathscr{H}_{\rm el}&=&-\sum_{\rm r}\int_{\varepsilon_{1-2}}\left[\frac{\alpha_1}{2\pi}(\varepsilon_1{+}\varepsilon_2){+}\frac{\alpha_2}{4\pi}(\varepsilon_1{+}\varepsilon_2)^2\right]b^{\dagger}_{\varepsilon_1 \rm r} b_{\varepsilon_2 \rm r},
\nonumber\\
\mathscr{H}_{\rm int}&{=}&\sum_{\rm r < r'}\int_{\varepsilon_{1-4}}\left[\frac{\phi_1}{\pi\nu}{+}\frac{\phi_2}{4\pi\nu}\sum^4_{\rm j=1}\varepsilon_{\rm j}\right]{:}b^{\dagger}_{\varepsilon_1 \rm r} b_{\varepsilon_2 \rm r} b^{\dagger}_{\varepsilon_3 \rm r'} b_{\varepsilon_4 \rm r'}{:}.\nonumber
\end{eqnarray}
The PH symmetric version of Eq.~\eqref{HamFL} is originally proposed by Nozieres~\cite{Nozieres} which is commonly known as Nozieres FL theory.
In Eq.~\eqref{HamFL} the density of states per species for a one dimensional channel is represented by the symbol $\nu$. The scattering (elastic) effects in the FL are accounted for by the Hamiltonian $\mathscr{H}_{\rm el}$, where $\alpha_1$ and $\alpha_2$ are the first and second generations of Nozieres FL coefficients respectively. The four fermions term represents the interaction part of the Hamiltonian $\mathscr{H}_{\rm int}$ which is expressed in terms of FL parameters $\phi_1$ and $\phi_2$. These FL parameters are related to the associated Kondo temperature of the corresponding SU($N$) impurity. The FL parameters characterizing the scattering effects are connected to those of interaction effects by the relation, $\alpha_1{=}(N{-}1)\phi_1$ and $\alpha_2{=}(N{-}1)\phi_2/4$. In addition the Bethe ansatz provides further link between $\alpha_1$ and $\alpha_2$~\cite{mora1, mora2},
\begin{equation}\label{sane8}
\mathscr{A}\equiv\frac{\alpha_2}{\alpha^2_1}=\frac{N-2}{N-1}\frac{\Gamma(1/N)\tan(\pi/N)}{\sqrt{\pi}\Gamma\left(\frac{1}{2}+\frac{1}{N}\right)}\cot\left[\frac{m\pi}{N}\right].
\end{equation}
Where $\Gamma(x)$ is the Euler's gamma-function. Therefore the low energy FL Hamiltonian Eq.~\eqref{HamFL} is completely specified by only one FL parameter, say $\alpha_1$. We make a connection of $\alpha_1$ with the corresponding Kondo temperature such that $T^{{\rm SU}(N)}_K{=}1/\alpha_1$, the $N$-dependence in FL parameters is implicit.
Note that we have retained upto the four fermions term in Eq.~\eqref{HamFL}, the higher-order terms produce the current correction beyond cubic order in applied bias and temperature gradient which is beyond the scope of present work.

It is then a straightforward procedure to proceed with the calculation of physical observables by treating the scattering Hamiltonian $\mathscr{H}_{\rm el}$ and interaction part $\mathscr{H}_{\rm int}$ perturbatively. However, in the spirit of Nozieres phenomenology, the scattering effects are fully described by an energy-dependent phase shift $\delta^{\rm el}_{\rm r}(\varepsilon)$. The Kondo singlet acts as the scatterer for
the incoming electrons from the leads. Outgoing and
incoming electrons are then differ from each other by the
elastic phase shift $\delta^{\rm el}_{\rm r}(\varepsilon)$. The Nozieres FL parameters $\alpha_1$ and $\alpha_2$ are the first and second order coefficients in the Taylor-series expansion of elastic phase shift. While the scattering effects are easily accounted for by the elastic phase-shift, the perturbative treatment of $\mathscr{H}_{\rm int}$ produces complicated self-energy diagrams. This complication can be simplified a bit by including the Hartree contribution of self-energy in the elastic phase shift~\cite{mora1, mora2}. Then the Taylor expansion of phase shift reads
\begin{align}\label{kaule1}
\delta_{\rm r}(\varepsilon)&=\delta_0+\alpha_1\varepsilon+\alpha_2\varepsilon^2-\sum_{\rm r'\neq r}\Bigg[\phi_1\int^{\infty}_{-\infty}d\varepsilon \delta n_{\rm r'}(\varepsilon)\nonumber\\
&+\frac{\phi_2}{2}\left(\varepsilon\int^{\infty}_{-\infty}d\varepsilon \delta n_{\rm r'}(\varepsilon)+\int^{\infty}_{-\infty}d\varepsilon \varepsilon \delta n_{\rm r'}(\varepsilon)\right)\Bigg].
\end{align}
Here the zero-energy phase shift of SU($N$) Kondo impurity with $m$ electrons is
\begin{equation}
\delta_0=\frac{m\pi}{N}.
\end{equation} 
In Eq.~\eqref{kaule1} we used the definition of the actual FL quasi-particle distribution relative to the Fermi-energy $\varepsilon_{\rm F}$ as $\delta n_{\rm r}(\varepsilon)\equiv n_{\rm r}(\varepsilon)-{\rm \Theta}(\varepsilon_{\rm F}-\varepsilon)=\langle b^{\dagger}_{k\rm r}b_{k\rm r}\rangle-{\rm \Theta}(\varepsilon_{\rm F}-\varepsilon)$, $\Theta$ being the step function. Using Eq.~\eqref{sane5} we expressed the average $\langle b^{\dagger}_{k\rm r}b_{k\rm r}\rangle$ in terms of the equilibrium Fermi-distribution functions $f_{\rm \gamma}(\varepsilon){=}\left[1+\exp\left(\frac{\varepsilon-\mu_{\gamma}}{T_{\gamma}}\right)\right]^{-1}$ of the left and right reservoirs; $\langle b^{\dagger}_{k\rm r}b_{k\rm r}\rangle=\cos^2\theta f_{\rm L}+\sin^2\theta f_{\rm R}$. In addition, we have implemented the specific choice of Fermi-level such that
\begin{equation}\label{kaule2}
\int^{\infty}_{-\infty}d\varepsilon\delta n_{\rm r}(\varepsilon)=0.
\end{equation}
This equation is always satisfied as far as the condition $\mu_{\rm L}\cos^2\theta +\mu_{\rm R}\sin^2\theta=\varepsilon_{\rm F}$ is full-filled. We then made the following specification for the chemical potentials of the reservoirs,
\begin{align}\label{aama1}
\mu_{\rm L} &=\phantom{-}e\Delta V\sin^2\theta\equiv\phantom{-}\frac{e\Delta V}{2}\left(1-\mathcal{C}\right),\\
\mu_{\rm R} &=-e\Delta V\cos^2\theta\equiv-\frac{e\Delta V}{2}\left(1+\mathcal{C}\right)\label{aamasa},
\end{align}
to make $\varepsilon_{\rm F}{=}0.$
It is also noted that the details related to the choice of the temperatures in the reservoirs do not affects the necessary condition to satisfy the Eq.~\eqref{kaule2}. To be more general, we do not yet impose any restriction on the choice of $T_{\rm L}$ and $T_{\rm R}$. Using these specification of chemical potentials and temperatures of the reservoirs, the straightforward integration of phase shift expression Eq.~\eqref{kaule1} leads
\begin{align}
\label{phase_shift}
&\delta_{\rm r} (\varepsilon)=\delta_0+\alpha_1\varepsilon+\alpha_2\left(\varepsilon^2- {\cal A}\right).
\end{align}
To obtained Eq.~\eqref{phase_shift} we have made the use of FL identity $\alpha_2=(N{-}1)\phi_2/4$ and the new definition,
\begin{equation}\nonumber
{\cal A}{=}\frac{1}{6}\left[(\pi T_{\rm L})^2(1{+}\mathcal{C}){+}(\pi T_{\rm R})^2(1{-}\mathcal{C}){+}\frac{3}{2}(1{-}\mathcal{C}^2)(e\Delta V)^2\right].
\end{equation}
In the following section the scattering effects in addition to the Hartree contribution to the self energy correction will be accounted for by the Eq.~\eqref{phase_shift}. To obtain the self energy correction beyond Hartree contribution we will be treating the interaction Hamiltonian $\mathscr{H}_{\rm int}$ perturbatively with the small parameters $\left(e\Delta V, T_{\rm L}, T_{\rm R}\right)/T^{{\rm SU}(N)}_K$.
\vspace*{-5mm}
\section{Current Calculation}\label{current}
\vspace*{-3mm}
Using the basis of scattering states that includes the elastic effects and Hartree term, we cast the charge current expression into the form~\cite{mora1, dee2}
\begin{equation}\label{kaule4}
\hat{I}{=}\frac{e}{2h\nu}\sum_{\rm r}\sin2\theta\left[a^{\dagger}_{\rm r}(x)b_{\rm r}(x){-}a^{\dagger}_{\rm r}(-x)\mathscr{E} b_{\rm r}(-x){+}{\rm H.c.}\right],
\end{equation}
for $b_{\rm r}(x){=}\sum_{k}b_{k\rm r}e^{ikx}$ and $\mathscr{E}b_{\rm r}(x){=}\sum_{k}\mathscr{E}_k b_{k\rm r}e^{ikx}$.
To write Eq.~\eqref{kaule4} we have also omitted the terms of the form $\sum_{\rm r, p{=}\pm}{\rm p}a^{\dagger}_{\rm r}({\rm p}x)a_{\rm r}({\rm p}x)$ since they do not produce finite contribution to the mean current. In addition, we expressed $N\times N$ scattering matrix $\mathscr{E}_{k}$ in terms of phase shift expression Eq.~\eqref{phase_shift} such that $\mathscr{E}_k{=}\exp[2i\delta_{\rm r}(\varepsilon_k)]$. To compute the various observables from Eq.~\eqref{kaule4} we need the following averages directly obtained from the Glazman-Raikh rotation
\begin{equation}\label{sane5}
 \left(%
\begin{array}{c}
  \langle b^{\dagger}_{k}b_{k }\rangle \\
 \langle a^{\dagger}_{k }a_{k }\rangle \\
 \langle b^{\dagger}_{k }a_{k}\rangle
\end{array}%
\right){=} \left(%
\begin{array}{ccc}
  \cos^2\theta & \phantom{-}\sin^2\theta & 0 \\
  \sin^2\theta & \phantom{-}\cos^2\theta  & 0 \\
   \frac{\sin2\theta}{2} & - \frac{\sin2\theta}{2} & 0 \\
\end{array}%
\right)\left(%
\begin{array}{c}
  f_{{\rm L}}(\varepsilon_k) \\
  f_{{\rm R}}(\varepsilon_k) \\
  0\\
\end{array}%
\right).
\end{equation}
The average of Eq.~\eqref{kaule4} provides the elastic current (including the corresponding Hartree contribution) which have the compact form analogous to the Landauer-B\"uttiker expression
\begin{equation}\label{LB_formula}
I_{\rm el}=\frac{e}{h}\sum^N_{\rm r}\int^{\infty}_{-\infty} d\varepsilon\; {\cal T}_{\rm r}(\varepsilon)\; \left[f_{\rm L}(\varepsilon)-f_{\rm R}(\varepsilon)\right].
\end{equation}
The effective transmission coefficient ${\cal T}_{\rm r}(\varepsilon)$ is completely specified by the phase shift expression Eq.~\eqref{phase_shift}; ${\cal T}_{{\rm r}}(\varepsilon){\equiv}\left(1-\mathcal{C}^2\right)\sin^2[\delta_{\rm r}(\varepsilon)]$. To write ${\cal T}_{\rm r}(\varepsilon)$ into more tractable form, we perform its Taylor expansion in energy and retained upto the second order terms,
\begin{align}\label{baa1}
{\cal T}_{\rm r}(\varepsilon)=\left(1-\mathcal{C}^2\right)&\Big[\mathcal{T}_0-\alpha_2\mathcal{A}\sin2\delta_0+\alpha_1\sin2\delta_0\;
\varepsilon\nonumber\\
& + \left(\alpha^2_1\cos2\delta_0+\alpha_2\sin2\delta_0\right)\varepsilon^2\Big].
\end{align}
Here $\mathcal{T}_0{=}\sin^2\delta_0$ is the zero energy transmission coefficient. Then it is trivial procedure to compute the elastic current by plugging in Eq.~\eqref{baa1} into Eq.~\eqref{LB_formula}. The exact computation of Eq.~\eqref{LB_formula} follows from the consideration of following integrals~\cite{dee2,dee3},
\begin{equation}\label{int}
\mathcal{K}_{\rm n}=\int^{\infty}_{-\infty} \varepsilon^{\rm n} \left[f_{\rm L}(\varepsilon)-f_{\rm R}(\varepsilon)\right]d\varepsilon,\;{\rm n}=0, 1\;\text{and}\;2.
\end{equation}
Conventional way of calculating the integrals in Eq.~\eqref{int} consists of Sommerfeld expansion of $\Delta f(\varepsilon)\equiv f_{\rm L}(\varepsilon)-f_{\rm R}(\varepsilon)$ in the small parameters $\Delta T{\equiv}T_{\rm L}{-}T_{\rm R}$ and $\Delta V$. However, the Fourier-transform technique allows us to compute Eq.~\eqref{int} exactly. Fourier transforming the function $\Delta f(\varepsilon)$ into real time reads,
\begin{equation}\label{int1}
\Delta f(t)=\frac{1}{2\pi}\int^{\infty}_{-\infty} d\varepsilon\;e^{-i\varepsilon t} \Delta f(\varepsilon).
\end{equation}
Performing the ${\rm n}$-times partial differentiation of Eq.~\eqref{int1} and taking the limit $t{\to}0$ we get,
\begin{equation}\label{inth}
\frac{2\pi}{(-i)^{\rm n}}\left. \frac{\partial^{\rm n}\Delta f(t)}{\partial t^{\rm n}}\right|_{t=0}=\int^{\infty}_{-\infty} d\varepsilon \varepsilon^{\rm n}\Delta f(\varepsilon).
\end{equation}
Fourier transformation of the Fermi distributions of left and right reservoirs allows us to write 
\begin{equation}\label{int2}
\Delta f(t)=\frac{i}{2}\left[\frac{T_{\rm L}e^{-i\mu_{\rm L}t}}{\sinh(\pi T_{\rm L} t)}-\frac{T_{\rm R}e^{-i\mu_{\rm R}t}}{\sinh(\pi T_{\rm R} t)}\right].
\end{equation}
Plugging in Eq.~\eqref{int2} into Eq.~\eqref{inth} with the chemical potentials as specified in Eqs.~\eqref{aama1} and~\eqref{aamasa} we obtain $\mathcal{K}_{\rm 0}{=}e\Delta V$, $\mathcal{K}_{\rm 1}{=}\left[(\pi T_{\rm L})^2{-}(\pi T_{\rm R})^2{-}3\mathcal{C} (e\Delta V)^2\right]/6$ and
\begin{align}\nonumber
\mathcal{K}_{\rm 2}{=}{\frac{e\Delta V}{3}}{{\Big[}\frac{(e\Delta V)^2}{4}\left(1{+}3\mathcal{C}^2\right){+}\frac{1{-}\mathcal{C}}{2}(\pi T_{\rm L})^2{+}\frac{1{+}\mathcal{C}}{2}(\pi T_{\rm R})^2{\Big]}}.
\end{align}
For completeness we re-express the elastic current in terms of the integrals in Eq.~\eqref{int} as
\begin{align}\label{kaule6}
I_{\rm el}=\frac{Ne\left(1-\mathcal{C}^2\right)}{h}&\Big[ \left(\mathcal{T}_0-\alpha_2\mathcal{A}\sin2\delta_0\right)\mathcal{K}_{\rm 0}
+\alpha_1\sin2\delta_0\mathcal{K}_{\rm 1}\nonumber\\
& +\left(\alpha^2_1\cos2\delta_0+\alpha_2\sin2\delta_0\right)\mathcal{K}_{\rm 2}\Big].
\end{align}
Now we turn to the discussion of inelastic effects leaving aside the Hartree contributions, which has been already accounted for by the phase shift expressed in Eq.~\eqref{phase_shift}. As we anticipated earlier, the perturbative treatment of $\mathscr{H}_{\rm int}$ imparts the interaction corrections to the charge current. This approach requires the expressions of non-interaction Green's functions (GFs) described by $\mathscr{H}_0$. \DK{The matrices of the non-interacting GFs in Keldysh space~\cite{keldysh} are given by }
\begin{align}\label{matrixgf}
\mathcal{G}_{bb/aa}(k,\varepsilon)&{=}\frac{1}{\varepsilon{-}\varepsilon_k}\tau_z{+} i\pi\begin{pmatrix}
{\rm F}_{b/a} &{\rm F}_{b/a}{+}1\\
{\rm F}_{b/a}{-}1& {\rm F}_{b/a}
\end{pmatrix}\delta(\varepsilon{-}\varepsilon_k),\nonumber\\
\mathcal{G}_{ba/ab}(k,\varepsilon)&=i\pi\begin{pmatrix}
1 &\phantom{-}1\\
1& \phantom{-}1
\end{pmatrix}{\rm F}_{ab}\;\delta(\varepsilon-\varepsilon_k).
\end{align}
Here the parameters ${\rm F}_{b/a}(\varepsilon)$ and ${\rm F}_{ab}(\varepsilon_k)$ are expressed in terms of different populations; ${\rm F}_b(\varepsilon_k){=}2\langle b^{\dagger}_k b_k\rangle{-}1$, ${\rm F}_a(\varepsilon_k){=}2\langle a^{\dagger}_k a_k\rangle{-}1$ and ${\rm F}_{ab}{=}2\langle b^{\dagger}_k a_k\rangle$. The z-component of Pauli-matrix is represented by $\tau_z$. However, in the flat-band limit only the off-diagonal parts of $\mathcal{G}_{bb}(k, \varepsilon)$, namely $\mathcal{G}^{+-}_{bb}(k, \varepsilon)$ and $\mathcal{G}^{-+}_{bb}(k, \varepsilon)$ produce the finite contribution to the charge current. The straightforward mathematical steps provide the following Fourier-transformed real-time GFs
\begin{align}
\mathcal{G}^{+-}_{bb}(t)&{=}{-}\frac{\pi\nu}{2}\left[\frac{T_{\rm L} (1{+}\mathcal{C}) e^{-i\mu_{\rm L}
t}}{\sinh(\pi T_{\rm L} t)}{+}\frac{T_{\rm R}(1{-}\mathcal{C}) e^{-i\mu_{\rm R} t}}{\sinh(\pi
T_{\rm R} t)}\right],\nonumber\\
\mathcal{G}_{ab/ba}(t)&{=}{-}\frac{\pi\nu}{2}\sqrt{1-\mathcal{C}^2}\left[\frac{T_{\rm L} e^{-i\mu_{\rm L}
t}}{\sinh(\pi T_{\rm L} t)}{-}\frac{T_{\rm R} e^{-i\mu_{\rm R} t}}{\sinh(\pi T_{\rm R} t)}\right].
\label{dec13}
\end{align}
Here $\mathcal{G}^{+-}_{bb}(t)$ and $\mathcal{G}^{-+}_{bb}(t)$ are connected by causality relations. In practice, the GFs expressed in Eqs.~\eqref{matrixgf} and~\eqref{dec13} are sufficient for the calculation of charge current.
To calculate the inelastic correction to the charge current we then apply the perturbation theory using Keldysh formalism \cite{keldysh},
\begin{equation}\label{wick}
\delta I_{\text{in}}=\langle T_C \hat{I}(t)e^{-i\int dt^{\prime} \mathscr{H}_{\text{int}}(t^{\prime})}\rangle,
\end{equation}
where $C$ denotes the double-side 
$\eta{=}\pm$ Keldysh contour and $T_C$ is corresponding time-ordering operator. We used the expression of charge current operator Eq.~\eqref{kaule4} and interaction Hamiltonian $\mathscr{H}_{\rm int}$ into Eq.~\eqref{wick} to obtain the interaction correction to the charge current
\begin{equation}\label{lc}
\delta I_{\rm in}=\mathscr{Z}\int^{\infty}_{-\infty} \frac{d\varepsilon}{2\pi}\left(\Sigma^{-+}-\Sigma^{+-}\right)(\varepsilon) i\pi\nu \Delta f(\varepsilon).
\end{equation}
To arrived from Eq.~\eqref{wick} to Eq.~\eqref{lc} we have already subtracted the diverging terms, which amounts the renormalization of FL coefficients (see Ref.~\cite{mora1} for detail). In addition, we introduced the new notation via; $\mathscr{Z}{=}\frac{N(N-1)}{h}e\pi\left(1-\mathcal{C}^2\right) \cos2\delta_0$. The self-energies in Eq.~\eqref{lc} are expressed in real-time as
\begin{align}\label{rama}
\Sigma^{\eta_1\eta_2}(t)=&\left(\frac{\phi_1}{\pi\nu^2}\right)^2\sum_{k_{1-3}}\Big[ \mathcal{G}^{\eta_1\eta_2}_{bb}(k_1, t) \nonumber\\
&\mathcal{G}^{\eta_2\eta_1}_{bb}(k_2, {-}t)
\mathcal{G}^{\eta_1 \eta_2}_{bb}(k_3, t)\Big].
\end{align}
To compute the self-energies, now we specify the temperatures of the left and right reservoirs $T_{\rm R}{=}T$ and $T_{\rm L}{=}T{+}\Delta T$ with $\Delta T{>}0$. In practice one can numerically solve for the self-energy using the GFs of Eq.~\eqref{dec13}. However, it is manageable to find the analytical expression of the self energy difference to the first order in $\Delta T$ and second order in $e\Delta V$ which reads
\begin{align}\label{lc1}
\left(\Sigma^{-+}{-}\Sigma^{+-}\right)(\varepsilon)=\frac{\phi^2_1}{i\pi\nu}
\Big[&\frac{3}{4}(e\Delta V)^2(1-\mathcal{C}^2)
+\varepsilon^2+(\pi T)^2\nonumber\\
& +\frac{\Delta T}{T}(\pi T)^2\left(1{+}\mathcal{C}\right)
\Big].
\end{align}
To arrive from Eq.~\eqref{rama} to Eq.~\eqref{lc1} we came across the integral of the form,
\begin{equation}\label{rama2}
\mathcal{Z}(a, T)=\int^{\infty}_{-\infty}\frac{e^{iat}}{\sinh^3(\pi T t)}\;dt.
\end{equation}
\DK{The singularity of the integral in Eq.~\eqref{rama2} is removed by shifting the time contour by $i\eta $, $\eta\to 0$ in the complex plane. The parameter $\eta$ is chosen such that $\eta D{\gg}1$ and $(\eta T, \eta \Delta T, \eta \Delta V)\ll 1$ with $D$ the band cutoff.
We chose the rectangular contour enclosing the singularity at $t{=}0$ and use the Cauchy's residue theorem to arrive at the result,}
\begin{equation}\label{rama3}
\mathcal{Z}(a, T)=-i\pi\;\frac{a^2+(\pi T)^2}{(\pi T)^2}\frac{1}{\exp(a/T)+1}.
\end{equation}
\DK{Equation~\eqref{lc1} contains all possible terms up to the linear response in $\Delta T$ and $\Delta V$. Therefore plugging in Eq.~\eqref{lc1} into Eq.~\eqref{lc} provides interaction correction up to the quadratic order in $\Delta T$ and $\Delta V$. To make interaction contributions to the charge current more symmetrical with that of elastic effects, we write}
\begin{align}\label{kaule10}
\delta I_{\rm in}=&\frac{Ne(1-\mathcal{C}^2)}{h}\frac{1}{2}\frac{1}{N-1}\cos2\delta_0\;\alpha^2_1\Big[\mathcal{K}_2\nonumber\\
&+\left(\frac{\Delta T}{T}(\pi T)^2\left(1+\mathcal{C}\right)+(\pi T)^2
\right)\mathcal{K}_0\Big].
\end{align}
This equation correctly reproduces the interaction correction up to the quadratic response with the coefficients $\mathcal{K}_{0, 2}$ given in Eq.~\eqref{int}. Using Eq.~\eqref{kaule6} and~\eqref{kaule10}, the charge current is given by
\begin{equation}
I_{\rm c}=I_{\rm el}+\delta I_{\rm int}.
\end{equation} 
\vspace*{-3mm}
\section{Results and discussion}\label{results}
\vspace*{-3mm}
The non-linear Seebeck effect is quantified by the Seebeck coefficient defined as the ratio of thermo-voltage developed under the condition of zero charge current, $\Delta V_{\rm th}\equiv\left.\Delta V\right|_{I_{\rm c}=0}$, to the applied temperature gradient~\cite{kimh,kim},
\begin{equation}\label{mann1}
\mathcal{S}\equiv \left.-\frac{\Delta V_{\rm th}}{T_{\rm L}- T_{\rm R}}\right|_{I_{\rm c}=0}.
\end{equation}
In fact, the Seebeck coefficient Eq.~\eqref{mann1} contains additional information than the electrical and thermal conductance measurements~\cite{ben}. While the electrical conductance depends merely on the density of states at the Fermi level, Seebeck coefficient reveals its slope~\cite{Zlatic}. In addition, the Seebeck coefficient provides the useful informations related to the average energy of charge carriers contributing to the transport processes~\cite{mat}. We characterize the Seebeck coefficient of a SU($N$) Kondo impurity by defining the dimensionless form of charge current accounting upto the quadratic responses,
\begin{align}\label{bhanji1}
&\mathcal{J}(N, m)\equiv\frac{I_{\rm c}(N, m)}{G_0(N)T^{\rm SU(N)}_K}\\
&=\mathscr{L}^1_{1}\Delta\overline{V}{+}\mathscr{L}^1_{2}\Delta\overline{T}
{+}\mathscr{L}^2_{1}\Delta\overline{V}^2{+}\mathscr{L}^2_{2}\Delta\overline{T}^2
{+}\mathscr{L}^{11}_{12}\Delta\overline{V}\Delta\overline{T}\nonumber.
\end{align}
The maximum conductance of SU($N$) Kondo impurity  in the presence of asymmetry is expressed by the relation $G_0(N)=\left(1-\mathcal{C}^2\right)Ne^2/h$. From now we use specify the electronic charge $e{=}{-}1$ and the convention $\Delta V{>}0$ and $\Delta T>0$. \DK{The quantities written in over-line letters represent that they are normalized with corresponding Kondo temperature: $\overline{T}\equiv T/T^{\rm SU(N)}_K$, $\Delta\overline{ T}\equiv \Delta T/T^{\rm SU(N)}_K$ and $\Delta \overline{V}\equiv \Delta V/T^{\rm SU(N)}_K$.} From Eq.~\eqref{kaule6} and~\eqref{kaule10} we obtained the transport coefficients $\mathscr{L}^i_j$ and $\mathscr{L}^{11}_{12}$, $i, j=1, 2$ for the SU($N$) Kondo impurity,
\begin{align}\label{bhanji2}
\mathscr{L}^1_1=&\left[\sin^2\left(\frac{ \pi  m}{N}\right)+\frac{1}{3} \frac{N+1}{N-1} \cos\left(\frac{2 \pi  m}{N}\right) (\pi\overline{T})^2\right],\nonumber\\
\mathscr{L}^1_2=&-\frac{\pi^2}{3} \overline{T} \sin \left(\frac{2 \pi  m}{N}\right),\;\;
\mathscr{L}^2_1=\frac{1}{2}\mathcal{C} \sin \left(\frac{2 \pi  m}{N}\right),\nonumber\\
\mathscr{L}^2_2=&-\frac{\pi^2}{6}\sin \left(\frac{2 \pi  m}{N}\right),\nonumber\\
\mathscr{L}^{11}_{12}=&-\frac{\pi^2}{3}\overline{T} \Big[ \mathscr{B} \cos \left(\frac{2 \pi  m}{N}\right)+2 \mathcal{C}\mathscr{A}\sin \left(\frac{2 \pi  m}{N}\right)\Big].
\end{align}
\DK{The coefficient $\mathscr{A}$ is defined in Eq.~\eqref{sane8} and $\mathscr{B}$ stands for}
\begin{align}\label{ram11}
\mathscr{B}\equiv\frac{\mathcal{C} (N-2)-N-1}{N-1}.
\end{align}
From Eq.~\eqref{bhanji2} it is seen that the transport coefficients accounting for the linear and quadratic correction in temperature gradient are connected by the relation $\mathscr{L}^1_2=2\overline{T}\mathscr{L}^2_2$. It is apparent that, merely the asymmetry of the junction is responsible to have the quadratic correction in voltage bias.  For half-filled SU($N$) Kondo effects, we observed that $\mathscr{L}^1_2=\mathscr{L}^2_1=\mathscr{L}^2_2=0$, therefore, corresponding thermoelectric properties are governed by only two coefficients $\mathscr{L}^1_1$ and $\mathscr{L}^{11}_{12}$. This fact explains that the  half-filled SU($N$) Kondo impurity do not offers finite thermo-power even in quadratic-response level of calculations. Another important conclusion can be drawn form Eq.~\eqref{bhanji2} is as follows; for the perfectly symmetrical quarter-filled SU($N$) Kondo correlated systems, the combine effects of temperature gradient and voltage bias tend to vanish $\left.\mathscr{L}^{11}_{12}\right|_{\mathcal{C}{=}0}(N, N/4){=}0$. Furthermore, the coefficients characterizing the voltage response do not acquire the temperature correction. These facts should make the non-linear thermoelectric measurement of beyond half-filled SU(4) systems as a trivial procedure. To have more insights of the thermoelectric production in SU($N$) Kondo systems, we solve the zero current condition of the Eq.~\eqref{bhanji1} to get the thermo-voltage upto the quadratic terms in $\Delta\overline{T}$, 
\begin{align}\label{add1}
-\Delta\overline{V}_{\rm th}&=\mathcal{S}^{\rm LR}\Delta \overline{T}+
\delta\mathcal{S}(\Delta\overline{T})^2+\mathcal{O}(\Delta\overline{T})^3.
\end{align}
The Seebeck coefficient $\mathcal{S}$ as defined in Eq.~\eqref{mann1} then takes the form,
\begin{equation}
\mathcal{S}=\mathcal{S}^{\rm LR}+\delta\mathcal{S}\Delta\overline{T}+\mathcal{O}(\Delta\overline{T})^2.
\end{equation}
Here $\mathcal{S}^{\rm LR}$ is the linear response Seebeck coefficient and its first order $\Delta\overline{T}$ correction is defined by $\delta\mathcal{S}$,
\begin{align}
\mathcal{S}^{\rm LR}&\equiv \frac{\mathscr{L}^1_2}{\mathscr{L}^1_1},\\
\delta\mathcal{S}&\equiv \left[\frac{\mathscr{L}^2_2}{\mathscr{L}^1_1}-\frac{\mathscr{L}^1_2\mathscr{L}^{11}_{12}}{\left(\mathscr{L}^1_1\right)^2}+\frac{\left(\mathscr{L}^1_2\right)^2\mathscr{L}^2_1}{\left(\mathscr{L}^1_1\right)^3}\right].\label{merobaa5}
\end{align}
The transport coefficients defining the linear response Seebeck coefficient $\mathcal{S}^{\rm LR}$ are independent of asymmetry parameter $\mathcal{C}$. However, the first order correction $\delta\mathcal{S}$ bears the strong dependences on the asymmetry parameter via the transport coefficients $\mathscr{L}^2_1$ and $\mathscr{L}^{11}_{12}$. In addition, for the symmetrical setups, we use the Eq.~\eqref{bhanji2} to express the correction factor $\delta\mathcal{S}$ entirely in terms of linear-response coefficients,
\begin{equation}\label{add2}
\left.\delta\mathcal{S}\right|_{\mathcal{C}{=}0}=\frac{\mathcal{S}^{\rm LR}}{\overline{T}}\left[\frac{\sin^2\left(\frac{ \pi  m}{N}\right)}{\mathscr{L}^1_1}-\frac{1}{2}\right].
\end{equation}
\DK{To study the effects of coupling asymmetry on the thermoelectric transport properties, we categorize the SU($N$) Kondo impurity into two broad classes, namely, half-filled (PH symmetric) and beyond half-filled and discuss them separately.}
\vspace*{-5mm}
\subsubsection{PH symmetric SU($N$) Kondo effects}
\vspace*{-3mm}
\begin{figure}[t]
\includegraphics[scale=0.62]{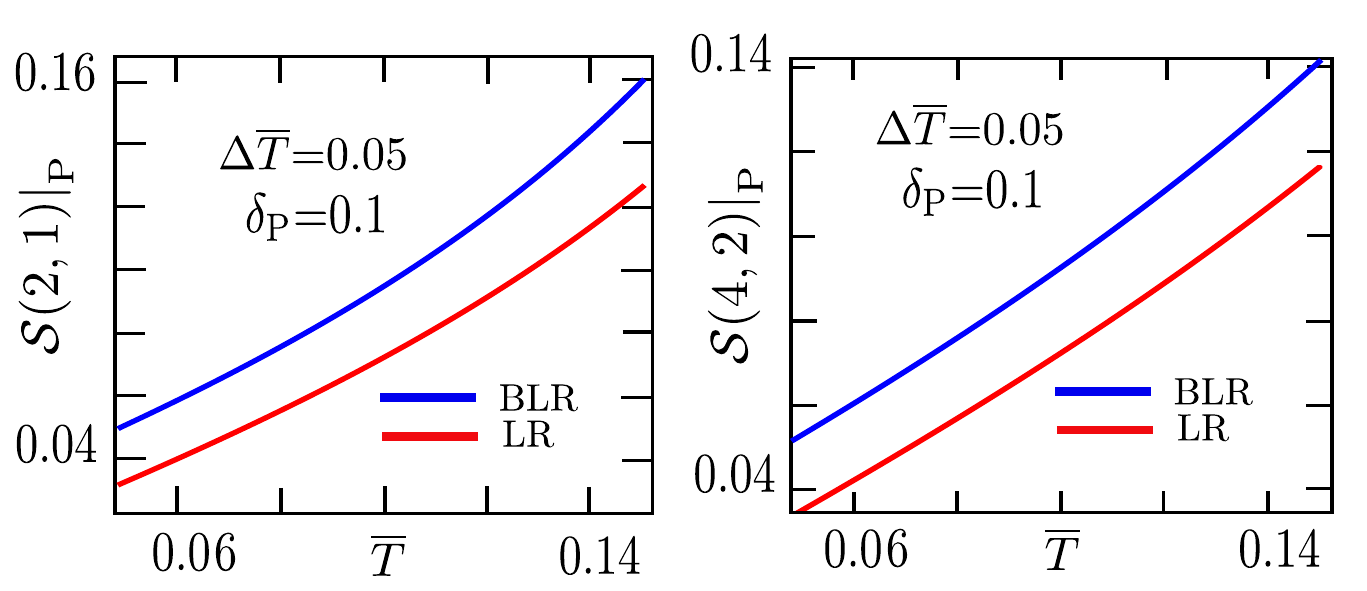}
\caption{Linear (LR) and non linear (BLR) Seebeck coefficients with PH symmetric SU(2) and SU(4) Kondo effects for fixed value of the potential scattering $\delta_{\rm P}$.}\label{yemk1}
\vspace*{-3mm}
\end{figure}
As we anticipated earlier that for the half-filled SU($N$) Kondo effects the transport coefficients satisfy the relation $\mathscr{L}^1_2{=}\mathscr{L}^2_1{=}\mathscr{L}^2_2{=}0$, therefore, corresponding thermoelectric properties are derived solely from the coefficients $\mathscr{L}^1_1$ and $\mathscr{L}^{11}_{12}$. The non-zero transport coefficients of PH symmetric SU($N$) Kondo effects are summarized below,
\begin{align}\label{add3}
\mathscr{L}^1_1(N, N/2)&=\left[1-\frac{1}{3} \frac{N+1}{N-1}(\pi\overline{T})^2\right],\\
\mathscr{L}^{11}_{12}(N, N/2)&=\frac{1}{3\overline{T}} \left[ \frac{\mathcal{C} (N-2)-N-1}{N-1}\right](\pi\overline{T})^2.
\end{align}
While for the conventional SU(2) Kondo effects the parameter $\mathcal{C}$ does not affects the cross-coefficient, the corresponding measurement in SU($N>2$) PH symmetric systems depends on the coupling asymmetry. However, the PH symmetry of the Kondo impurity realized in QDs is exact only if the dot is tuned to the middle of Coulomb valley~\cite{ carmi1}. This indicates the possibility of breaking the underlying PH symmetry. This weakly broken PH symmetry of Kondo correlated systems is accounted for by re-normalizing the reference phase shifts such that~\cite{lea,Pust, jvd},
\begin{equation}\label{merobaa}
\delta_0\to\tilde{\delta_0}=\delta_0+\delta_{\rm P},\;\;\delta_0\gg\delta_{\rm P}.
\end{equation} 
This potential scattering provides the repulsive
interactions which breaks the Kondo singlet and contributes to inelastic processes~\cite{yg}. The first order transport coefficients in Eq.~\eqref{bhanji2} for PH symmetric Kondo correlated systems with an account of the potential scattering effects are then given by
\begin{align}\label{merobaa1}
\left.\mathscr{L}^1_1(N, N/2)\right|_{\rm P}&=\cos^2\delta_{\rm P}\left[1{-}\frac{(\pi\overline{T})^2}{3}\frac{N{+}1}{N{-}1}\frac{2\cos2\delta_{\rm P}}{1{+}\cos2\delta_{\rm P}}\right],\nonumber\\
\left.\mathscr{L}^1_2(N, N/2)\right|_{\rm P}&=\cos^2\delta_{\rm P}\left[\frac{(\pi\overline{T})^2}{3\overline{T}}\frac{2\sin2\delta_{\rm P}}{1+\cos2\delta_{\rm P}}\right].
\end{align}
Eq.~\eqref{merobaa1} allows us to compute the linear response Seebeck coefficient of PH symmetric SU($N$) Kondo effects with small potential scattering,
\begin{equation}\label{merobaa2}
\left.\mathcal{S}^{\rm LR}(N, N/2)\right|_{\rm P}=\frac{2}{3}\;\frac{1}{\overline{T}}\;\frac{(\pi\overline{T})^2}{1-\frac{(\pi\overline{T})^2}{3}\frac{N+1}{N-1}}\delta_{\rm P}+\mathcal{O}(\delta_{\rm P})^3.
\end{equation}
Note that due to the numerical factor $(N{+}1)/(N{-}1)$ in the denominator of Eq.~\eqref{merobaa2}, among PH symmetric generalizations of SU($N$) the SU(2) Kondo correlated systems offer highest value of the linear response Seebeck coefficient in the presence of finite potential scattering. Plugging in the Eq.~\eqref{merobaa} into the transport coefficients Eq.~\eqref{bhanji2} and using them into Eq.~\eqref{merobaa5}, we get the first order correction to the Seebeck coefficient upto the linear order in $\delta_{\rm P}$,
\begin{equation}\label{merobaa7}
\left.\delta\mathcal{S}(N, N/2)\right|_{\rm P}=\frac{\pi^2}{3}\;\frac{1{-}\frac{(\pi \overline{T})^2}{3}\left(\frac{N{+}1}{N{-}1}{+}2\mathscr{B}\right)}{\left[1{-}\frac{(\pi \overline{T})^2}{3}\frac{N{+}1}{N{-}1}\right]^2}\delta_{\rm P}{+}\mathcal{O}(\delta_{\rm P})^3.
\end{equation}
For SU(2) Kondo effects the correction Eq.~\eqref{merobaa7} is independent of the asymmetry parameter as can be inferred from Eq.~\eqref{ram11}. However for SU(4) and other PH symmetric version of SU($N$), the first order correction to the Seebeck effect is weakly asymmetry dependent via the coefficient $\mathscr{B}(\mathcal{C})$. The linear and non linear Seebeck coefficient with PH symmetric SU(2) and SU(4) Kondo effects are shown in Fig.~\ref{yemk1} with the choice  of potential scattering term $\delta_{\rm P}=0.1$ and temperature gradient $\Delta\overline{T}=0.05$. These significant enhancement of BLR Seebeck coefficients with respect to the corresponding LR contribution get further improve at relatively high reference temperature and large temperature drop across the junction.
\vspace*{-3mm}
\subsubsection{Beyond half-filled SU($N$) Kondo effects}
\vspace*{-3mm}
The Kondo correlated systems with $N{>}2$ provide the realization of paradigmatic PH-asymmetric setups. First we start form the SU(3) Kondo effects. The SU(3) Kondo effect can occur either with single electron or two electrons. Furthermore, the SU(3) Kondo systems do not offer the PH symmetric analog. The physics of SU(3) Kondo effect with one and two electrons is related with each other by PH symmetry transformation. Therefore, we discuss the single electron SU(3) Kondo systems, which will ultimately provide the corresponding informations of two electron case. For single electron SU(3) Kondo effects the transport coefficients in Eq.~\eqref{bhanji2} are simplified as 
\begin{align}\label{bhanji7}
\mathscr{L}^1_1(3, 1)&=\frac{3}{4}\left[1-\frac{4}{9}(\pi\overline{T})^2\right],\;\mathscr{L}^1_2(3, 1)=-\frac{\pi^2\overline{T}}{2\sqrt{3}},\nonumber\\
\mathscr{L}^2_1(3, 1)&=\mathcal{C}\frac{\sqrt{3}}{4},\;\;\mathscr{L}^2_2(3, 1)=-\frac{\pi^2}{4\sqrt{3}},\nonumber\\
\mathscr{L}^{11}_{12}(3,1)&=-\frac{\pi^2}{3}\overline{T} \left[\frac{\mathcal{C}}{4}\left(1-2\sqrt{\frac{3}{\pi}} \frac{\Gamma[1/3]}{\Gamma[5/6]}\right)-1\right].
\end{align}
Therefore while the cross coefficient $\mathscr{L}^{11}_{12}(3,1)$ is weakly asymmetry dependent, the coefficient $\mathscr{L}^2_1(3, 1)$ is strongly influenced by $\mathcal{C}$.
Since all the transport coefficients in Eq.~\eqref{bhanji7} are non-zero, one can solve the zero-current equation to get the thermo-voltage developed in SU(3) Kondo effects.

\begin{figure}
\includegraphics[scale=1.2]{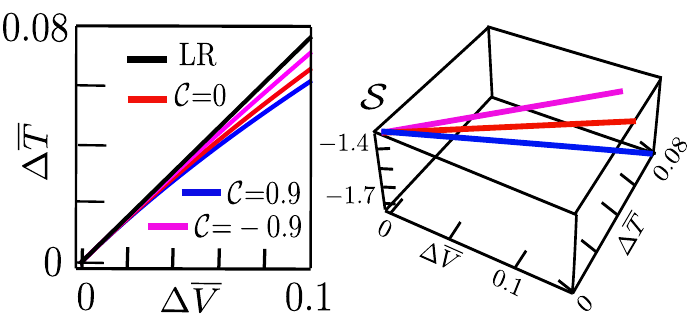}
\caption{Left panel: Plot of asymmetry dependent zero current lines in PH asymmetric SU(4) Kondo effects within the qudratic response level of calculations as a function of applied voltage bias and temperature gradient at reference temperature $\overline{T}{=}0.2$. Right panel: Corresponding Seebeck coefficients for given asymmetry parameter.}\label{mb}
\vspace*{-4mm}
\end{figure}

Now we turn to the discussion of SU(4) Kondo effects out of PH symmetric situation. The SU(4) Kondo effects can accommodate upto three electrons. While the two electron case is suffers from the PH symmetry, the single and three electron SU(4) systems are regarded to have good thermoelectric performance. Furthermore, the single electron and three electron systems are related to each other by the PH symmetry transformation. Therefore we discuss in details about the thermoelectric of single electron SU(4) Kondo effects. The corresponding transport coefficients are obtained as
\begin{align}\label{bhanji8}
\mathscr{L}^1_1(4, 1)&=\frac{1}{2},\;\;\mathscr{L}^1_2(4, 1)=-\frac{\pi^2}{3}\overline{T},\;\;\mathscr{L}^2_1(4, 1)=\frac{\mathcal{C}}{2},\nonumber\\
\mathscr{L}^2_2(4, 1)&=-\frac{\pi^2}{6},\;\;\mathscr{L}^{11}_{12}(4, 1)=-\frac{4 \pi ^{2} \mathcal{C} \overline{T}}{9\sqrt{\pi}}\frac{\Gamma[1/4]}{\Gamma[3/4]}.
\end{align}
The cross coefficient $\mathscr{L}^{11}_{12}(4, 1){\simeq}{-}7.32\mathcal{C}\overline{T}$ is very large as compared to other coefficients for relatively large asymmetry parameter. In addition the other coefficient $\mathscr{L}^2_1(4, 1)$ is also strongly asymmetry dependent. Presence of these coefficients is solely manifested by the finite asymmetry of the junction. Therefore we argue that measuring this cross coefficient would be useful while identifying the asymmetry of the junction in addition to its physical implications. Just form the structure of Eq.~\eqref{bhanji8}, it is seen that the thermoelectric transport properties of beyond half-filled SU(4) Kondo effects can be easily manipulated by tuning the junction asymmetry. It appears that the effect of asymmetry becomes more pronounce in relatively high temperature gradient regime. The asymmetry parameter $\mathcal{C}$ mainly causes to shift the zero-current line either upward or downward with respect to the perfectly symmetric setup. As shown in Fig.~\ref{mb} the positive value of the asymmetry parameter increases the thermo-voltage, while the opposite effects are apparent for the corresponding negative values. In addition, the beyond linear response contribution always overshoots the corresponding linear response value irrespective of the coupling asymmetry.
\vspace*{-5mm}
\subsubsection{Paradigmatic SU(4) Kondo effects}
\begin{figure}[t]
\includegraphics[scale=0.21]{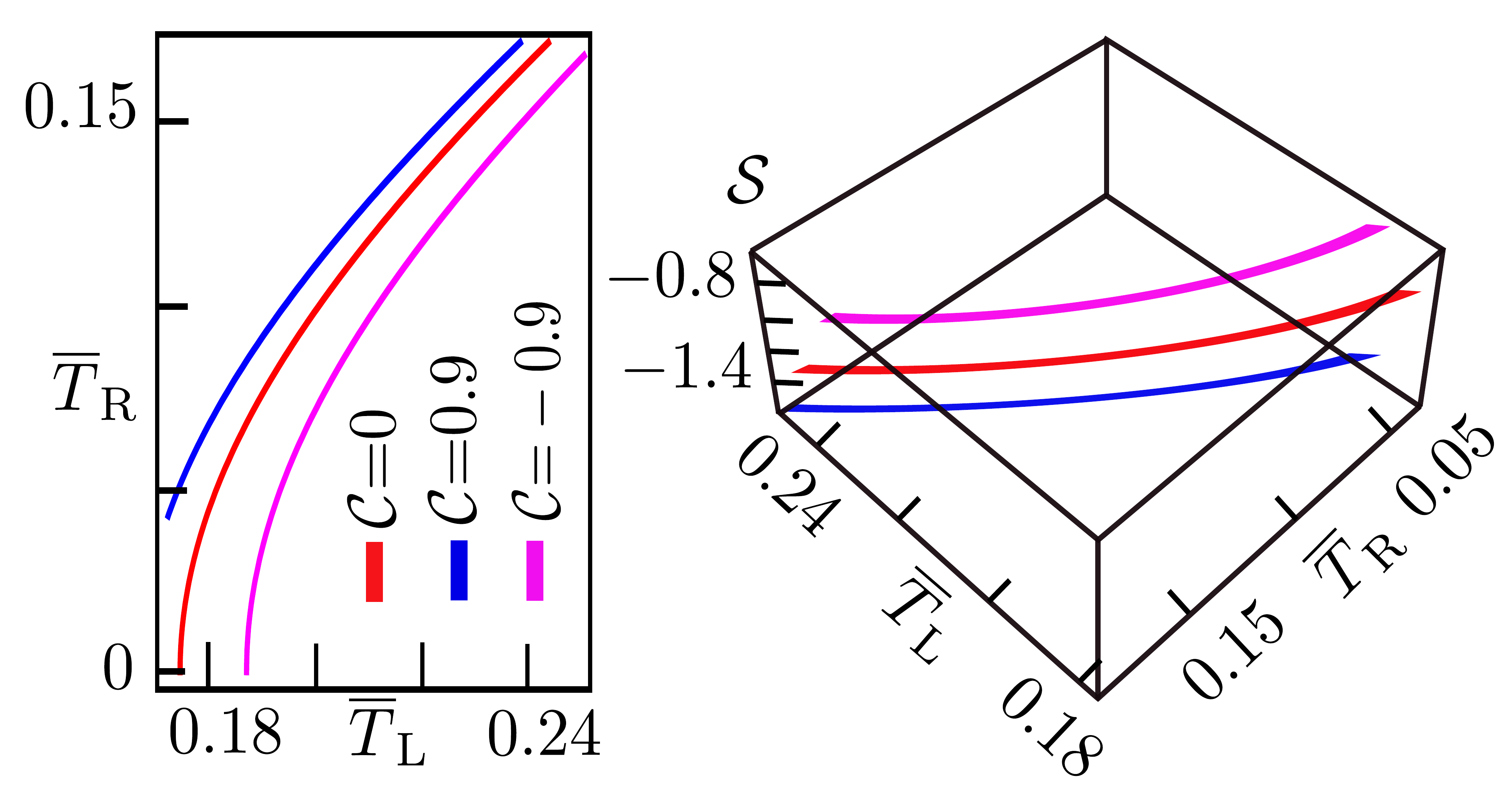}
\caption{Left panel: Lines of zero charge currents in a single electron SU(4) Kondo impurity within cubic response level of calculations. The temperatures of left and right reservoirs (normalized with corresponding Kondo temperature) are varied for given asymmetry parameter at fixed voltage drop $\Delta\overline{V}{=}0.1$. Right panel: The Seebeck coefficients as a function of asymmetry parameter with beyond half-filled SU(4) Kondo effects for fixed voltage drop $\Delta\overline{V}{=}0.1$.}\label{myy}
\vspace{-3mm}
\end{figure}

\vspace*{-3mm}
The cosine factor $\cos2\delta_0$ in front of the expression of the inelastic current dramatically modifies the low energy transport behavior of SU($N$) Kondo effects. In case of the SU($N$) systems with $m$ electrons satisfying the specific combination such that $m/N{=}(2{\rm n}+1)/4\;{\rm for}\;{\rm n=0\;{\rm and}\;1}$, the cosine factor $\cos2\delta_0$ in Eq.~\eqref{kaule10} amounts to nullify the whole expression. For these specific systems, the beyond Hartree contribution to the self-energy becomes zero being corresponding Hartree contribution finite. In addition, for PH symmetric SU(N) Kondo effects the Hartree contribution vanishes and beyond Hartree contribution becomes finite. Interestingly, the PH asymmetric SU(4) Kondo correlated systems offer vanishing non-Hartree contribution to the self-energy. Since the Hartree contributions can be straightforwardly accounted for by including it in phase shift, the beyond-half filled SU(4) systems can be exactly solved within cubic response and even beyond. This paradigmatic simplication is also applicable for some SU(12) generalizations. From Eq.~\eqref{kaule6} we obtained two non-zero cubic response coefficients $\mathscr{L}^3_1(4, 1)$ and $\mathscr{L}^{12}_{12}(4, 1)$ contributing to the charge current of beyond-half filled SU(4) Kondo impurity as
\begin{equation}\label{bhanji9}\nonumber
\left.\mathcal{J}(4, 1)\right|_{\rm cubic}=\mathscr{L}^3_1(4, 1)(\Delta\overline{V})^3+\mathscr{L}^{12}_{12}(4, 1)\Delta\overline{V}(\Delta\overline{T})^2.
\end{equation}
Here the transport coefficients are
\begin{align}\nonumber
\mathscr{L}^3_1(4, 1)&{=}{-}\frac{1{-}3\mathcal{C}^2}{9\sqrt{\pi}}\frac{\Gamma[1/4]}{\Gamma[3/4]},\;\mathscr{L}^{12}_{12}(4, 1){=}{-}\mathcal{C}\frac{2\pi^2}{9\sqrt{\pi}}\frac{\Gamma[1/4]}{\Gamma[3/4]}.
\end{align}
These equations show that for the perfectly symmetrical single electron SU(4) Kondo setups, the effects of voltage bias and temperature gradient are not correlated even in cubic response level of calculations. Therefore, only the asymmetry can derives these systems to have combine interplay of voltage bias and temperature gradient. The effects of asymmetry parameter on the Seebeck coefficient in cubic response level of calculations has been presented in Fig.~\ref{myy} with an example of single electron SU(4) Kondo effects. From Fig.~\ref{myy} it is seen that with the proper choice (positive value) of asymmetry parameter $\mathcal{C}$ the non-linear Seebeck coefficient gets significantly enhanced over the corresponding perfectly symmetrical coupling. \DK{This effect is associated with strong asymmetry of the beyond linear response transmission coefficient Eq.~\eqref{baa1}.}

Finally we want to mention that the non-linearly has been also studied by generalizing the definition of Seebeck coefficient with constant current condition~\cite{sabina, aligia,ue} such that
\begin{equation}\label{bhanji4}
\overline{\mathcal{S}}(N, m)=  \frac{\partial \mathcal{J}(N, m)}{\partial\Delta\overline{T}}\bigg/ \frac{\partial \mathcal{J}(N, m)}{\partial\Delta\overline{V}}.
\end{equation}
In the linear response level of calculation the response coefficient defined in Eq.~\eqref{bhanji4} coincides with the Seebeck coefficient given by Eq.~\eqref{mann1}. Though their behaviors in non-linear regime is quite different, it has been argued that the coefficient $\overline{\mathcal{S}}$ is indeed experimentally accessible~\cite{sabina} and can provide an important ingredient for the propose of temperature sensing. These effects have been already studied in conventional SU(2) Kondo regime accounting for the linear response of temperature gradient and finite voltage bias~\cite{sabina, aligia}. The central result of the paper expressed in Eq.~\eqref{bhanji2} paved a straightforward way of extending their study with an account of strong non-linearity in more exotic Kondo correlated system.
\vspace*{-4mm}
\section{Conclusions}\label{conclusion}
\vspace*{-2mm}
We developed a theoretical framework based on a local Fermi-liquid theory in combination with the out of equilibrium Keldysh approach to study the influences of coupling asymmetry on the thermoelectric transport of a strongly coupled SU($N$) Kondo impurity. \DK{While the linear response Seebeck coefficient is independent of coupling asymmetry, the fundamental role of non-linearity towards the enhancement of the Seebeck coefficient with a SU($N$) Kondo setup is explored.} In addition, we reported the great enhancement of Seebeck coefficient of Kondo impurities by proper tailoring the coupling asymmetry. We explore the importance of potential scattering on the thermoelectric characterization of PH symmetric SU($N$) Kondo effects. The presented analytical expressions of asymmetry dependent transport coefficients for general SU($N$) Kondo effects allow us to make a close connection of our findings with the experimentally studied SU(2) and SU(4) Kondo effects in complex QDs nano structures. Application of developed theoretical framework for the investigation of thermoelectric properties of more exotic Kondo problems such as multi-stage and multi-terminal Kondo screening appears to be a valid avenues for \DK{future} research.
\section*{Acknowledgement}
We thank Eva Andrei, Yigal Meir, Yuval Oreg, Giulio Casati and Giuliano Benenti for the illuminating discussions. This work was performed in part at Aspen Center for Physics, which is supported by National Science Foundation grant PHY-1607611. This work was
partially supported (MK) by a grant from the Simons Foundation.

\begin{thebibliography}{66}%
\makeatletter
\providecommand \@ifxundefined [1]{%
 \@ifx{#1\undefined}
}%
\providecommand \@ifnum [1]{%
 \ifnum #1\expandafter \@firstoftwo
 \else \expandafter \@secondoftwo
 \fi
}%
\providecommand \@ifx [1]{%
 \ifx #1\expandafter \@firstoftwo
 \else \expandafter \@secondoftwo
 \fi
}%
\providecommand \natexlab [1]{#1}%
\providecommand \enquote  [1]{``#1''}%
\providecommand \bibnamefont  [1]{#1}%
\providecommand \bibfnamefont [1]{#1}%
\providecommand \citenamefont [1]{#1}%
\providecommand \href@noop [0]{\@secondoftwo}%
\providecommand \href [0]{\begingroup \@sanitize@url \@href}%
\providecommand \@href[1]{\@@startlink{#1}\@@href}%
\providecommand \@@href[1]{\endgroup#1\@@endlink}%
\providecommand \@sanitize@url [0]{\catcode `\\12\catcode `\$12\catcode
  `\&12\catcode `\#12\catcode `\^12\catcode `\_12\catcode `\%12\relax}%
\providecommand \@@startlink[1]{}%
\providecommand \@@endlink[0]{}%
\providecommand \url  [0]{\begingroup\@sanitize@url \@url }%
\providecommand \@url [1]{\endgroup\@href {#1}{\urlprefix }}%
\providecommand \urlprefix  [0]{URL }%
\providecommand \Eprint [0]{\href }%
\providecommand \doibase [0]{http://dx.doi.org/}%
\providecommand \selectlanguage [0]{\@gobble}%
\providecommand \bibinfo  [0]{\@secondoftwo}%
\providecommand \bibfield  [0]{\@secondoftwo}%
\providecommand \translation [1]{[#1]}%
\providecommand \BibitemOpen [0]{}%
\providecommand \bibitemStop [0]{}%
\providecommand \bibitemNoStop [0]{.\EOS\space}%
\providecommand \EOS [0]{\spacefactor3000\relax}%
\providecommand \BibitemShut  [1]{\csname bibitem#1\endcsname}%
\let\auto@bib@innerbib\@empty
\bibitem [{\citenamefont {Dresselhaus}\ \emph {et~al.}(1999)\citenamefont
  {Dresselhaus}, \citenamefont {Dresselhaus}, \citenamefont {Sun},
  \citenamefont {Zhang}, \citenamefont {Cronin},\ and\ \citenamefont
  {Koga}}]{ld0}%
  \BibitemOpen
  \bibfield  {author} {\bibinfo {author} {\bibfnamefont {M.~S.}\ \bibnamefont
  {Dresselhaus}}, \bibinfo {author} {\bibfnamefont {G.}~\bibnamefont
  {Dresselhaus}}, \bibinfo {author} {\bibfnamefont {X.}~\bibnamefont {Sun}},
  \bibinfo {author} {\bibfnamefont {Z.}~\bibnamefont {Zhang}}, \bibinfo
  {author} {\bibfnamefont {S.~B.}\ \bibnamefont {Cronin}}, \ and\ \bibinfo
  {author} {\bibfnamefont {T.}~\bibnamefont {Koga}},\ }\href {\doibase
  10.1134/1.1130849} {\bibfield  {journal} {\bibinfo  {journal} {Physics of the
  Solid State}\ }\textbf {\bibinfo {volume} {41}},\ \bibinfo {pages} {679}
  (\bibinfo {year} {1999})}\BibitemShut {NoStop}%
\bibitem [{\citenamefont {Datta}(1995)}]{ld00}%
  \BibitemOpen
  \bibfield  {author} {\bibinfo {author} {\bibfnamefont {S.}~\bibnamefont
  {Datta}},\ }\href {\doibase 10.1017/CBO9780511805776} {\emph {\bibinfo
  {title} {Electronic Transport in Mesoscopic Systems}}},\ Cambridge Studies in
  Semiconductor Physics and Microelectronic Engineering\ (\bibinfo  {publisher}
  {Cambridge University Press},\ \bibinfo {year} {1995})\BibitemShut {NoStop}%
\bibitem [{\citenamefont {Dresselhaus}\ \emph {et~al.}(2007)\citenamefont
  {Dresselhaus}, \citenamefont {Chen}, \citenamefont {Tang}, \citenamefont
  {Yang}, \citenamefont {Lee}, \citenamefont {Wang}, \citenamefont {Ren},
  \citenamefont {Fleurial},\ and\ \citenamefont {Gogna}}]{ld1}%
  \BibitemOpen
  \bibfield  {author} {\bibinfo {author} {\bibfnamefont {M.}~\bibnamefont
  {Dresselhaus}}, \bibinfo {author} {\bibfnamefont {G.}~\bibnamefont {Chen}},
  \bibinfo {author} {\bibfnamefont {M.}~\bibnamefont {Tang}}, \bibinfo {author}
  {\bibfnamefont {R.}~\bibnamefont {Yang}}, \bibinfo {author} {\bibfnamefont
  {H.}~\bibnamefont {Lee}}, \bibinfo {author} {\bibfnamefont {D.}~\bibnamefont
  {Wang}}, \bibinfo {author} {\bibfnamefont {Z.}~\bibnamefont {Ren}}, \bibinfo
  {author} {\bibfnamefont {J.-P.}\ \bibnamefont {Fleurial}}, \ and\ \bibinfo
  {author} {\bibfnamefont {P.}~\bibnamefont {Gogna}},\ }\href {\doibase
  10.1002/adma.200600527} {\bibfield  {journal} {\bibinfo  {journal} {Advanced
  Materials}\ }\textbf {\bibinfo {volume} {19}},\ \bibinfo {pages} {1043}
  (\bibinfo {year} {2007})}\BibitemShut {NoStop}%
\bibitem [{\citenamefont {Benenti}\ \emph {et~al.}(2017)\citenamefont
  {Benenti}, \citenamefont {Casati}, \citenamefont {Saito},\ and\ \citenamefont
  {Whitney}}]{casti}%
  \BibitemOpen
  \bibfield  {author} {\bibinfo {author} {\bibfnamefont {G.}~\bibnamefont
  {Benenti}}, \bibinfo {author} {\bibfnamefont {G.}~\bibnamefont {Casati}},
  \bibinfo {author} {\bibfnamefont {K.}~\bibnamefont {Saito}}, \ and\ \bibinfo
  {author} {\bibfnamefont {R.}~\bibnamefont {Whitney}},\ }\href {\doibase
  https://doi.org/10.1016/j.physrep.2017.05.008} {\bibfield  {journal}
  {\bibinfo  {journal} {Physics Reports}\ }\textbf {\bibinfo {volume} {694}},\
  \bibinfo {pages} {1 } (\bibinfo {year} {2017})},\ \bibinfo {note}
  {fundamental aspects of steady-state conversion of heat to work at the
  nanoscale}\BibitemShut {NoStop}%
\bibitem [{\citenamefont {Blanter}\ and\ \citenamefont
  {Nazarov}(2009)}]{Blanter}%
  \BibitemOpen
  \bibfield  {author} {\bibinfo {author} {\bibfnamefont {Y.~M.}\ \bibnamefont
  {Blanter}}\ and\ \bibinfo {author} {\bibfnamefont {Y.~V.}\ \bibnamefont
  {Nazarov}},\ }\href@noop {} {\emph {\bibinfo {title} {Quantum Transport:
  Introduction to Nanoscience}}}\ (\bibinfo  {publisher} {Cambridge University
  Press, Cambridge, England},\ \bibinfo {year} {2009})\BibitemShut {NoStop}%
\bibitem [{\citenamefont {Zhang}\ and\ \citenamefont {Zhao}(2015)}]{Z}%
  \BibitemOpen
  \bibfield  {author} {\bibinfo {author} {\bibfnamefont {X.}~\bibnamefont
  {Zhang}}\ and\ \bibinfo {author} {\bibfnamefont {L.-D.}\ \bibnamefont
  {Zhao}},\ }\href {\doibase https://doi.org/10.1016/j.jmat.2015.01.001}
  {\bibfield  {journal} {\bibinfo  {journal} {Journal of Materiomics}\ }\textbf
  {\bibinfo {volume} {1}},\ \bibinfo {pages} {92 } (\bibinfo {year}
  {2015})}\BibitemShut {NoStop}%
\bibitem [{\citenamefont {Costi}\ and\ \citenamefont {Zlatic}(2010)}]{costi1}%
  \BibitemOpen
  \bibfield  {author} {\bibinfo {author} {\bibfnamefont {T.~A.}\ \bibnamefont
  {Costi}}\ and\ \bibinfo {author} {\bibfnamefont {V.}~\bibnamefont {Zlatic}},\
  }\href {\doibase 10.1103/PhysRevB.81.235127} {\bibfield  {journal} {\bibinfo
  {journal} {Phys. Rev. B}\ }\textbf {\bibinfo {volume} {81}},\ \bibinfo
  {pages} {235127} (\bibinfo {year} {2010})}\BibitemShut {NoStop}%
\bibitem [{\citenamefont {Kondo}(1964)}]{kondo}%
  \BibitemOpen
  \bibfield  {author} {\bibinfo {author} {\bibfnamefont {J.}~\bibnamefont
  {Kondo}},\ }\href {\doibase 10.1143/PTP.32.37} {\bibfield  {journal}
  {\bibinfo  {journal} {Progress of Theoretical Physics}\ }\textbf {\bibinfo
  {volume} {32}},\ \bibinfo {pages} {37} (\bibinfo {year} {1964})}\BibitemShut
  {NoStop}%
\bibitem [{\citenamefont {Scheibner}\ \emph {et~al.}(2005)\citenamefont
  {Scheibner}, \citenamefont {Buhmann}, \citenamefont {Reuter}, \citenamefont
  {Kiselev},\ and\ \citenamefont {Molenkamp}}]{kiselev}%
  \BibitemOpen
  \bibfield  {author} {\bibinfo {author} {\bibfnamefont {R.}~\bibnamefont
  {Scheibner}}, \bibinfo {author} {\bibfnamefont {H.}~\bibnamefont {Buhmann}},
  \bibinfo {author} {\bibfnamefont {D.}~\bibnamefont {Reuter}}, \bibinfo
  {author} {\bibfnamefont {M.~N.}\ \bibnamefont {Kiselev}}, \ and\ \bibinfo
  {author} {\bibfnamefont {L.~W.}\ \bibnamefont {Molenkamp}},\ }\href {\doibase
  10.1103/PhysRevLett.95.176602} {\bibfield  {journal} {\bibinfo  {journal}
  {Phys. Rev. Lett.}\ }\textbf {\bibinfo {volume} {95}},\ \bibinfo {pages}
  {176602} (\bibinfo {year} {2005})}\BibitemShut {NoStop}%
\bibitem [{\citenamefont {Jezouin}\ \emph {et~al.}(2013)\citenamefont
  {Jezouin}, \citenamefont {Parmentier}, \citenamefont {Anthore}, \citenamefont
  {Gennser}, \citenamefont {Cavanna}, \citenamefont {Jin},\ and\ \citenamefont
  {Pierre}}]{if3}%
  \BibitemOpen
  \bibfield  {author} {\bibinfo {author} {\bibfnamefont {S.}~\bibnamefont
  {Jezouin}}, \bibinfo {author} {\bibfnamefont {F.~D.}\ \bibnamefont
  {Parmentier}}, \bibinfo {author} {\bibfnamefont {A.}~\bibnamefont {Anthore}},
  \bibinfo {author} {\bibfnamefont {U.}~\bibnamefont {Gennser}}, \bibinfo
  {author} {\bibfnamefont {A.}~\bibnamefont {Cavanna}}, \bibinfo {author}
  {\bibfnamefont {Y.}~\bibnamefont {Jin}}, \ and\ \bibinfo {author}
  {\bibfnamefont {F.}~\bibnamefont {Pierre}},\ }\href {\doibase
  10.1126/science.1241912} {\bibfield  {journal} {\bibinfo  {journal}
  {Science}\ }\textbf {\bibinfo {volume} {342}},\ \bibinfo {pages} {601}
  (\bibinfo {year} {2013})}\BibitemShut {NoStop}%
\bibitem [{\citenamefont {Iftikhar}\ \emph {et~al.}(2015)\citenamefont
  {Iftikhar}, \citenamefont {Jezouin}, \citenamefont {Anthore}, \citenamefont
  {Gennser}, \citenamefont {Parmentier}, \citenamefont {Cavanna},\ and\
  \citenamefont {Pierre}}]{if1}%
  \BibitemOpen
  \bibfield  {author} {\bibinfo {author} {\bibfnamefont {Z.}~\bibnamefont
  {Iftikhar}}, \bibinfo {author} {\bibfnamefont {S.}~\bibnamefont {Jezouin}},
  \bibinfo {author} {\bibfnamefont {A.}~\bibnamefont {Anthore}}, \bibinfo
  {author} {\bibfnamefont {U.}~\bibnamefont {Gennser}}, \bibinfo {author}
  {\bibfnamefont {F.~D.}\ \bibnamefont {Parmentier}}, \bibinfo {author}
  {\bibfnamefont {A.}~\bibnamefont {Cavanna}}, \ and\ \bibinfo {author}
  {\bibfnamefont {F.}~\bibnamefont {Pierre}},\ }\href
  {http://dx.doi.org/10.1038/nature15384} {\bibfield  {journal} {\bibinfo
  {journal} {Nature}\ }\textbf {\bibinfo {volume} {526}},\ \bibinfo {pages}
  {233} (\bibinfo {year} {2015})}\BibitemShut {NoStop}%
\bibitem [{\citenamefont {Jezouin}\ \emph {et~al.}(2016)\citenamefont
  {Jezouin}, \citenamefont {Iftikhar}, \citenamefont {Anthore}, \citenamefont
  {Parmentier}, \citenamefont {Gennser}, \citenamefont {Cavanna}, \citenamefont
  {Ouerghi}, \citenamefont {Levkivskyi}, \citenamefont {Idrisov}, \citenamefont
  {Sukhorukov}, \citenamefont {Glazman},\ and\ \citenamefont {Pierre}}]{if2}%
  \BibitemOpen
  \bibfield  {author} {\bibinfo {author} {\bibfnamefont {S.}~\bibnamefont
  {Jezouin}}, \bibinfo {author} {\bibfnamefont {Z.}~\bibnamefont {Iftikhar}},
  \bibinfo {author} {\bibfnamefont {A.}~\bibnamefont {Anthore}}, \bibinfo
  {author} {\bibfnamefont {F.~D.}\ \bibnamefont {Parmentier}}, \bibinfo
  {author} {\bibfnamefont {U.}~\bibnamefont {Gennser}}, \bibinfo {author}
  {\bibfnamefont {A.}~\bibnamefont {Cavanna}}, \bibinfo {author} {\bibfnamefont
  {A.}~\bibnamefont {Ouerghi}}, \bibinfo {author} {\bibfnamefont {I.~P.}\
  \bibnamefont {Levkivskyi}}, \bibinfo {author} {\bibfnamefont
  {E.}~\bibnamefont {Idrisov}}, \bibinfo {author} {\bibfnamefont {E.~V.}\
  \bibnamefont {Sukhorukov}}, \bibinfo {author} {\bibfnamefont {L.~I.}\
  \bibnamefont {Glazman}}, \ and\ \bibinfo {author} {\bibfnamefont
  {F.}~\bibnamefont {Pierre}},\ }\href {http://dx.doi.org/10.1038/nature19072}
  {\bibfield  {journal} {\bibinfo  {journal} {Nature}\ }\textbf {\bibinfo
  {volume} {536}},\ \bibinfo {pages} {60} (\bibinfo {year} {2016})}\BibitemShut
  {NoStop}%
\bibitem [{\citenamefont {Ferrier}\ \emph {et~al.}(2016)\citenamefont
  {Ferrier}, \citenamefont {Arakawa}, \citenamefont {Hata}, \citenamefont
  {Fujiwara}, \citenamefont {Delagrange}, \citenamefont {Weil}, \citenamefont
  {Deblock}, \citenamefont {Sakano}, \citenamefont {Oguri},\ and\ \citenamefont
  {Kobayashi}}]{if4}%
  \BibitemOpen
  \bibfield  {author} {\bibinfo {author} {\bibfnamefont {M.}~\bibnamefont
  {Ferrier}}, \bibinfo {author} {\bibfnamefont {T.}~\bibnamefont {Arakawa}},
  \bibinfo {author} {\bibfnamefont {T.}~\bibnamefont {Hata}}, \bibinfo {author}
  {\bibfnamefont {R.}~\bibnamefont {Fujiwara}}, \bibinfo {author}
  {\bibfnamefont {R.}~\bibnamefont {Delagrange}}, \bibinfo {author}
  {\bibfnamefont {R.}~\bibnamefont {Weil}}, \bibinfo {author} {\bibfnamefont
  {R.}~\bibnamefont {Deblock}}, \bibinfo {author} {\bibfnamefont
  {R.}~\bibnamefont {Sakano}}, \bibinfo {author} {\bibfnamefont
  {A.}~\bibnamefont {Oguri}}, \ and\ \bibinfo {author} {\bibfnamefont
  {K.}~\bibnamefont {Kobayashi}},\ }\href {http://dx.doi.org/10.1038/nphys3556}
  {\bibfield  {journal} {\bibinfo  {journal} {Nature Physics}\ }\textbf
  {\bibinfo {volume} {12}},\ \bibinfo {pages} {230} (\bibinfo {year}
  {2016})}\BibitemShut {NoStop}%
\bibitem [{\citenamefont {Svilans}\ \emph {et~al.}(2018)\citenamefont
  {Svilans}, \citenamefont {Josefsson}, \citenamefont {Burke}, \citenamefont
  {Fahlvik}, \citenamefont {Thelander}, \citenamefont {Linke},\ and\
  \citenamefont {Leijnse}}]{heiner}%
  \BibitemOpen
  \bibfield  {author} {\bibinfo {author} {\bibfnamefont {A.}~\bibnamefont
  {Svilans}}, \bibinfo {author} {\bibfnamefont {M.}~\bibnamefont {Josefsson}},
  \bibinfo {author} {\bibfnamefont {A.~M.}\ \bibnamefont {Burke}}, \bibinfo
  {author} {\bibfnamefont {S.}~\bibnamefont {Fahlvik}}, \bibinfo {author}
  {\bibfnamefont {C.}~\bibnamefont {Thelander}}, \bibinfo {author}
  {\bibfnamefont {H.}~\bibnamefont {Linke}}, \ and\ \bibinfo {author}
  {\bibfnamefont {M.}~\bibnamefont {Leijnse}},\ }\href {\doibase
  10.1103/PhysRevLett.121.206801} {\bibfield  {journal} {\bibinfo  {journal}
  {Phys. Rev. Lett.}\ }\textbf {\bibinfo {volume} {121}},\ \bibinfo {pages}
  {206801} (\bibinfo {year} {2018})}\BibitemShut {NoStop}%
\bibitem [{\citenamefont {Dutta}\ \emph {et~al.}(2019)\citenamefont {Dutta},
  \citenamefont {Majidi}, \citenamefont {Garcia~Corral}, \citenamefont
  {Erdman}, \citenamefont {Florens}, \citenamefont {Costi}, \citenamefont
  {Courtois},\ and\ \citenamefont {Winkelmann}}]{paulo}%
  \BibitemOpen
  \bibfield  {author} {\bibinfo {author} {\bibfnamefont {B.}~\bibnamefont
  {Dutta}}, \bibinfo {author} {\bibfnamefont {D.}~\bibnamefont {Majidi}},
  \bibinfo {author} {\bibfnamefont {A.}~\bibnamefont {Garcia~Corral}}, \bibinfo
  {author} {\bibfnamefont {P.~A.}\ \bibnamefont {Erdman}}, \bibinfo {author}
  {\bibfnamefont {S.}~\bibnamefont {Florens}}, \bibinfo {author} {\bibfnamefont
  {T.~A.}\ \bibnamefont {Costi}}, \bibinfo {author} {\bibfnamefont
  {H.}~\bibnamefont {Courtois}}, \ and\ \bibinfo {author} {\bibfnamefont
  {C.~B.}\ \bibnamefont {Winkelmann}},\ }\href {\doibase
  10.1021/acs.nanolett.8b04398} {\bibfield  {journal} {\bibinfo  {journal}
  {Nano Letters}\ }\textbf {\bibinfo {volume} {19}},\ \bibinfo {pages} {506}
  (\bibinfo {year} {2019})}\BibitemShut {NoStop}%
\bibitem [{\citenamefont {Karki}\ and\ \citenamefont {Kiselev}(2017)}]{dee1}%
  \BibitemOpen
  \bibfield  {author} {\bibinfo {author} {\bibfnamefont {D.~B.}\ \bibnamefont
  {Karki}}\ and\ \bibinfo {author} {\bibfnamefont {M.~N.}\ \bibnamefont
  {Kiselev}},\ }\href {\doibase 10.1103/PhysRevB.96.121403} {\bibfield
  {journal} {\bibinfo  {journal} {Phys. Rev. B}\ }\textbf {\bibinfo {volume}
  {96}},\ \bibinfo {pages} {121403(R)} (\bibinfo {year} {2017})}\BibitemShut
  {NoStop}%
\bibitem [{\citenamefont {Azema}\ \emph {et~al.}(2012)\citenamefont {Azema},
  \citenamefont {Dar\'e}, \citenamefont {Sch\"afer},\ and\ \citenamefont
  {Lombardo}}]{azema}%
  \BibitemOpen
  \bibfield  {author} {\bibinfo {author} {\bibfnamefont {J.}~\bibnamefont
  {Azema}}, \bibinfo {author} {\bibfnamefont {A.-M.}\ \bibnamefont {Dar\'e}},
  \bibinfo {author} {\bibfnamefont {S.}~\bibnamefont {Sch\"afer}}, \ and\
  \bibinfo {author} {\bibfnamefont {P.}~\bibnamefont {Lombardo}},\ }\href
  {\doibase 10.1103/PhysRevB.86.075303} {\bibfield  {journal} {\bibinfo
  {journal} {Phys. Rev. B}\ }\textbf {\bibinfo {volume} {86}},\ \bibinfo
  {pages} {075303} (\bibinfo {year} {2012})}\BibitemShut {NoStop}%
\bibitem [{\citenamefont {{Karki}}\ and\ \citenamefont
  {{Kiselev}}(2019)}]{dee4}%
  \BibitemOpen
  \bibfield  {author} {\bibinfo {author} {\bibfnamefont {D.~B.}\ \bibnamefont
  {{Karki}}}\ and\ \bibinfo {author} {\bibfnamefont {M.~N.}\ \bibnamefont
  {{Kiselev}}},\ }\href@noop {} {\bibfield  {journal} {\bibinfo  {journal}
  {arXiv e-prints}\ ,\ \bibinfo {pages} {arXiv:1906.00724}} (\bibinfo {year}
  {2019})}\BibitemShut {NoStop}%
\bibitem [{\citenamefont {Jarillo-Herrero}\ \emph {et~al.}(2005)\citenamefont
  {Jarillo-Herrero}, \citenamefont {Kong}, \citenamefont {van~der Zant},
  \citenamefont {Dekker}, \citenamefont {Kouwenhoven},\ and\ \citenamefont
  {Franceschi}}]{jh}%
  \BibitemOpen
  \bibfield  {author} {\bibinfo {author} {\bibfnamefont {P.}~\bibnamefont
  {Jarillo-Herrero}}, \bibinfo {author} {\bibfnamefont {J.}~\bibnamefont
  {Kong}}, \bibinfo {author} {\bibfnamefont {H.~S.}\ \bibnamefont {van~der
  Zant}}, \bibinfo {author} {\bibfnamefont {C.}~\bibnamefont {Dekker}},
  \bibinfo {author} {\bibfnamefont {L.~P.}\ \bibnamefont {Kouwenhoven}}, \ and\
  \bibinfo {author} {\bibfnamefont {S.~D.}\ \bibnamefont {Franceschi}},\ }\href
  {https://doi.org/10.1038/nature03422} {\bibfield  {journal} {\bibinfo
  {journal} {Nature}\ }\textbf {\bibinfo {volume} {434}},\ \bibinfo {pages}
  {484} (\bibinfo {year} {2005})}\BibitemShut {NoStop}%
\bibitem [{\citenamefont {Makarovski}\ \emph
  {et~al.}(2007{\natexlab{a}})\citenamefont {Makarovski}, \citenamefont {Liu},\
  and\ \citenamefont {Finkelstein}}]{sasa0}%
  \BibitemOpen
  \bibfield  {author} {\bibinfo {author} {\bibfnamefont {A.}~\bibnamefont
  {Makarovski}}, \bibinfo {author} {\bibfnamefont {J.}~\bibnamefont {Liu}}, \
  and\ \bibinfo {author} {\bibfnamefont {G.}~\bibnamefont {Finkelstein}},\
  }\href {\doibase 10.1103/PhysRevLett.99.066801} {\bibfield  {journal}
  {\bibinfo  {journal} {Phys. Rev. Lett.}\ }\textbf {\bibinfo {volume} {99}},\
  \bibinfo {pages} {066801} (\bibinfo {year} {2007}{\natexlab{a}})}\BibitemShut
  {NoStop}%
\bibitem [{\citenamefont {Makarovski}\ \emph
  {et~al.}(2007{\natexlab{b}})\citenamefont {Makarovski}, \citenamefont
  {Zhukov}, \citenamefont {Liu},\ and\ \citenamefont {Finkelstein}}]{sasa}%
  \BibitemOpen
  \bibfield  {author} {\bibinfo {author} {\bibfnamefont {A.}~\bibnamefont
  {Makarovski}}, \bibinfo {author} {\bibfnamefont {A.}~\bibnamefont {Zhukov}},
  \bibinfo {author} {\bibfnamefont {J.}~\bibnamefont {Liu}}, \ and\ \bibinfo
  {author} {\bibfnamefont {G.}~\bibnamefont {Finkelstein}},\ }\href {\doibase
  10.1103/PhysRevB.75.241407} {\bibfield  {journal} {\bibinfo  {journal} {Phys.
  Rev. B}\ }\textbf {\bibinfo {volume} {75}},\ \bibinfo {pages} {241407}
  (\bibinfo {year} {2007}{\natexlab{b}})}\BibitemShut {NoStop}%
\bibitem [{\citenamefont {Ferrier}\ \emph {et~al.}(2017)\citenamefont
  {Ferrier}, \citenamefont {Arakawa}, \citenamefont {Hata}, \citenamefont
  {Fujiwara}, \citenamefont {Delagrange}, \citenamefont {Deblock},
  \citenamefont {Teratani}, \citenamefont {Sakano}, \citenamefont {Oguri},\
  and\ \citenamefont {Kobayashi}}]{ll}%
  \BibitemOpen
  \bibfield  {author} {\bibinfo {author} {\bibfnamefont {M.}~\bibnamefont
  {Ferrier}}, \bibinfo {author} {\bibfnamefont {T.}~\bibnamefont {Arakawa}},
  \bibinfo {author} {\bibfnamefont {T.}~\bibnamefont {Hata}}, \bibinfo {author}
  {\bibfnamefont {R.}~\bibnamefont {Fujiwara}}, \bibinfo {author}
  {\bibfnamefont {R.}~\bibnamefont {Delagrange}}, \bibinfo {author}
  {\bibfnamefont {R.}~\bibnamefont {Deblock}}, \bibinfo {author} {\bibfnamefont
  {Y.}~\bibnamefont {Teratani}}, \bibinfo {author} {\bibfnamefont
  {R.}~\bibnamefont {Sakano}}, \bibinfo {author} {\bibfnamefont
  {A.}~\bibnamefont {Oguri}}, \ and\ \bibinfo {author} {\bibfnamefont
  {K.}~\bibnamefont {Kobayashi}},\ }\href {\doibase
  10.1103/PhysRevLett.118.196803} {\bibfield  {journal} {\bibinfo  {journal}
  {Phys. Rev. Lett.}\ }\textbf {\bibinfo {volume} {118}},\ \bibinfo {pages}
  {196803} (\bibinfo {year} {2017})}\BibitemShut {NoStop}%
\bibitem [{\citenamefont {Hata}\ \emph {et~al.}(2018)\citenamefont {Hata},
  \citenamefont {Delagrange}, \citenamefont {Arakawa}, \citenamefont {Lee},
  \citenamefont {Deblock}, \citenamefont {Bouchiat}, \citenamefont
  {Kobayashi},\ and\ \citenamefont {Ferrier}}]{ll1}%
  \BibitemOpen
  \bibfield  {author} {\bibinfo {author} {\bibfnamefont {T.}~\bibnamefont
  {Hata}}, \bibinfo {author} {\bibfnamefont {R.}~\bibnamefont {Delagrange}},
  \bibinfo {author} {\bibfnamefont {T.}~\bibnamefont {Arakawa}}, \bibinfo
  {author} {\bibfnamefont {S.}~\bibnamefont {Lee}}, \bibinfo {author}
  {\bibfnamefont {R.}~\bibnamefont {Deblock}}, \bibinfo {author} {\bibfnamefont
  {H.}~\bibnamefont {Bouchiat}}, \bibinfo {author} {\bibfnamefont
  {K.}~\bibnamefont {Kobayashi}}, \ and\ \bibinfo {author} {\bibfnamefont
  {M.}~\bibnamefont {Ferrier}},\ }\href {\doibase
  10.1103/PhysRevLett.121.247703} {\bibfield  {journal} {\bibinfo  {journal}
  {Phys. Rev. Lett.}\ }\textbf {\bibinfo {volume} {121}},\ \bibinfo {pages}
  {247703} (\bibinfo {year} {2018})}\BibitemShut {NoStop}%
\bibitem [{\citenamefont {Keller}\ \emph {et~al.}(2014)\citenamefont {Keller},
  \citenamefont {Amasha}, \citenamefont {Weymann}, \citenamefont {Moca},
  \citenamefont {Rau}, \citenamefont {Katine}, \citenamefont {Shtrikman},
  \citenamefont {Zar{\'a}nd},\ and\ \citenamefont {Goldhaber-Gordon}}]{keller}%
  \BibitemOpen
  \bibfield  {author} {\bibinfo {author} {\bibfnamefont {A.~J.}\ \bibnamefont
  {Keller}}, \bibinfo {author} {\bibfnamefont {S.}~\bibnamefont {Amasha}},
  \bibinfo {author} {\bibfnamefont {I.}~\bibnamefont {Weymann}}, \bibinfo
  {author} {\bibfnamefont {C.~P.}\ \bibnamefont {Moca}}, \bibinfo {author}
  {\bibfnamefont {I.~G.}\ \bibnamefont {Rau}}, \bibinfo {author} {\bibfnamefont
  {J.~A.}\ \bibnamefont {Katine}}, \bibinfo {author} {\bibfnamefont
  {H.}~\bibnamefont {Shtrikman}}, \bibinfo {author} {\bibfnamefont
  {G.}~\bibnamefont {Zar{\'a}nd}}, \ and\ \bibinfo {author} {\bibfnamefont
  {D.}~\bibnamefont {Goldhaber-Gordon}},\ }\href
  {http://dx.doi.org/10.1038/nphys2844} {\bibfield  {journal} {\bibinfo
  {journal} {Nature Physics}\ }\textbf {\bibinfo {volume} {10}},\ \bibinfo
  {pages} {145} (\bibinfo {year} {2014})}\BibitemShut {NoStop}%
\bibitem [{\citenamefont {Tettamanzi}\ \emph {et~al.}(2012)\citenamefont
  {Tettamanzi}, \citenamefont {Verduijn}, \citenamefont {Lansbergen},
  \citenamefont {Blaauboer}, \citenamefont {Calder\'on}, \citenamefont
  {Aguado},\ and\ \citenamefont {Rogge}}]{st}%
  \BibitemOpen
  \bibfield  {author} {\bibinfo {author} {\bibfnamefont {G.~C.}\ \bibnamefont
  {Tettamanzi}}, \bibinfo {author} {\bibfnamefont {J.}~\bibnamefont
  {Verduijn}}, \bibinfo {author} {\bibfnamefont {G.~P.}\ \bibnamefont
  {Lansbergen}}, \bibinfo {author} {\bibfnamefont {M.}~\bibnamefont
  {Blaauboer}}, \bibinfo {author} {\bibfnamefont {M.~J.}\ \bibnamefont
  {Calder\'on}}, \bibinfo {author} {\bibfnamefont {R.}~\bibnamefont {Aguado}},
  \ and\ \bibinfo {author} {\bibfnamefont {S.}~\bibnamefont {Rogge}},\ }\href
  {\doibase 10.1103/PhysRevLett.108.046803} {\bibfield  {journal} {\bibinfo
  {journal} {Phys. Rev. Lett.}\ }\textbf {\bibinfo {volume} {108}},\ \bibinfo
  {pages} {046803} (\bibinfo {year} {2012})}\BibitemShut {NoStop}%
\bibitem [{\citenamefont {Le~Hur}\ \emph {et~al.}(2007)\citenamefont {Le~Hur},
  \citenamefont {Simon},\ and\ \citenamefont {Loss}}]{hur}%
  \BibitemOpen
  \bibfield  {author} {\bibinfo {author} {\bibfnamefont {K.}~\bibnamefont
  {Le~Hur}}, \bibinfo {author} {\bibfnamefont {P.}~\bibnamefont {Simon}}, \
  and\ \bibinfo {author} {\bibfnamefont {D.}~\bibnamefont {Loss}},\ }\href
  {\doibase 10.1103/PhysRevB.75.035332} {\bibfield  {journal} {\bibinfo
  {journal} {Phys. Rev. B}\ }\textbf {\bibinfo {volume} {75}},\ \bibinfo
  {pages} {035332} (\bibinfo {year} {2007})}\BibitemShut {NoStop}%
\bibitem [{\citenamefont {Choi}\ \emph {et~al.}(2005)\citenamefont {Choi},
  \citenamefont {L\'opez},\ and\ \citenamefont {Aguado}}]{su4_21}%
  \BibitemOpen
  \bibfield  {author} {\bibinfo {author} {\bibfnamefont {M.-S.}\ \bibnamefont
  {Choi}}, \bibinfo {author} {\bibfnamefont {R.}~\bibnamefont {L\'opez}}, \
  and\ \bibinfo {author} {\bibfnamefont {R.}~\bibnamefont {Aguado}},\ }\href
  {\doibase 10.1103/PhysRevLett.95.067204} {\bibfield  {journal} {\bibinfo
  {journal} {Phys. Rev. Lett.}\ }\textbf {\bibinfo {volume} {95}},\ \bibinfo
  {pages} {067204} (\bibinfo {year} {2005})}\BibitemShut {NoStop}%
\bibitem [{\citenamefont {Eto}(2005)}]{su4_22}%
  \BibitemOpen
  \bibfield  {author} {\bibinfo {author} {\bibfnamefont {M.}~\bibnamefont
  {Eto}},\ }\href {\doibase 10.1143/JPSJ.74.95} {\bibfield  {journal} {\bibinfo
   {journal} {Journal of the Physical Society of Japan}\ }\textbf {\bibinfo
  {volume} {74}},\ \bibinfo {pages} {95} (\bibinfo {year} {2005})}\BibitemShut
  {NoStop}%
\bibitem [{\citenamefont {Lim}\ \emph {et~al.}(2006)\citenamefont {Lim},
  \citenamefont {Choi}, \citenamefont {Choi}, \citenamefont {L\'opez},\ and\
  \citenamefont {Aguado}}]{su4_23}%
  \BibitemOpen
  \bibfield  {author} {\bibinfo {author} {\bibfnamefont {J.~S.}\ \bibnamefont
  {Lim}}, \bibinfo {author} {\bibfnamefont {M.-S.}\ \bibnamefont {Choi}},
  \bibinfo {author} {\bibfnamefont {M.~Y.}\ \bibnamefont {Choi}}, \bibinfo
  {author} {\bibfnamefont {R.}~\bibnamefont {L\'opez}}, \ and\ \bibinfo
  {author} {\bibfnamefont {R.}~\bibnamefont {Aguado}},\ }\href {\doibase
  10.1103/PhysRevB.74.205119} {\bibfield  {journal} {\bibinfo  {journal} {Phys.
  Rev. B}\ }\textbf {\bibinfo {volume} {74}},\ \bibinfo {pages} {205119}
  (\bibinfo {year} {2006})}\BibitemShut {NoStop}%
\bibitem [{\citenamefont {Lim}\ \emph {et~al.}(2014)\citenamefont {Lim},
  \citenamefont {L{\'{o}}pez},\ and\ \citenamefont {S{\'{a}}nchez}}]{Lim}%
  \BibitemOpen
  \bibfield  {author} {\bibinfo {author} {\bibfnamefont {J.~S.}\ \bibnamefont
  {Lim}}, \bibinfo {author} {\bibfnamefont {R.}~\bibnamefont {L{\'{o}}pez}}, \
  and\ \bibinfo {author} {\bibfnamefont {D.}~\bibnamefont {S{\'{a}}nchez}},\
  }\href {\doibase 10.1088/1367-2630/16/1/015003} {\bibfield  {journal}
  {\bibinfo  {journal} {New Journal of Physics}\ }\textbf {\bibinfo {volume}
  {16}},\ \bibinfo {pages} {015003} (\bibinfo {year} {2014})}\BibitemShut
  {NoStop}%
\bibitem [{\citenamefont {Kleeorin}\ and\ \citenamefont {Meir}(2017)}]{su4_11}%
  \BibitemOpen
  \bibfield  {author} {\bibinfo {author} {\bibfnamefont {Y.}~\bibnamefont
  {Kleeorin}}\ and\ \bibinfo {author} {\bibfnamefont {Y.}~\bibnamefont
  {Meir}},\ }\href {\doibase 10.1103/PhysRevB.96.045118} {\bibfield  {journal}
  {\bibinfo  {journal} {Phys. Rev. B}\ }\textbf {\bibinfo {volume} {96}},\
  \bibinfo {pages} {045118} (\bibinfo {year} {2017})}\BibitemShut {NoStop}%
\bibitem [{\citenamefont {Carmi}\ \emph {et~al.}(2011)\citenamefont {Carmi},
  \citenamefont {Oreg},\ and\ \citenamefont {Berkooz}}]{aash}%
  \BibitemOpen
  \bibfield  {author} {\bibinfo {author} {\bibfnamefont {A.}~\bibnamefont
  {Carmi}}, \bibinfo {author} {\bibfnamefont {Y.}~\bibnamefont {Oreg}}, \ and\
  \bibinfo {author} {\bibfnamefont {M.}~\bibnamefont {Berkooz}},\ }\href
  {\doibase 10.1103/PhysRevLett.106.106401} {\bibfield  {journal} {\bibinfo
  {journal} {Phys. Rev. Lett.}\ }\textbf {\bibinfo {volume} {106}},\ \bibinfo
  {pages} {106401} (\bibinfo {year} {2011})}\BibitemShut {NoStop}%
\bibitem [{\citenamefont {L\'opez}\ \emph {et~al.}(2013)\citenamefont
  {L\'opez}, \citenamefont {Rejec}, \citenamefont {Martinek},\ and\
  \citenamefont {\ifmmode~\check{Z}\else \v{Z}\fi{}itko}}]{rok1}%
  \BibitemOpen
  \bibfield  {author} {\bibinfo {author} {\bibfnamefont {R.}~\bibnamefont
  {L\'opez}}, \bibinfo {author} {\bibfnamefont {T.~c.~v.}\ \bibnamefont
  {Rejec}}, \bibinfo {author} {\bibfnamefont {J.}~\bibnamefont {Martinek}}, \
  and\ \bibinfo {author} {\bibfnamefont {R.}~\bibnamefont
  {\ifmmode~\check{Z}\else \v{Z}\fi{}itko}},\ }\href {\doibase
  10.1103/PhysRevB.87.035135} {\bibfield  {journal} {\bibinfo  {journal} {Phys.
  Rev. B}\ }\textbf {\bibinfo {volume} {87}},\ \bibinfo {pages} {035135}
  (\bibinfo {year} {2013})}\BibitemShut {NoStop}%
\bibitem [{\citenamefont {Kita}\ \emph {et~al.}(2008)\citenamefont {Kita},
  \citenamefont {Sakano}, \citenamefont {Ohashi},\ and\ \citenamefont
  {Suga}}]{kita}%
  \BibitemOpen
  \bibfield  {author} {\bibinfo {author} {\bibfnamefont {T.}~\bibnamefont
  {Kita}}, \bibinfo {author} {\bibfnamefont {R.}~\bibnamefont {Sakano}},
  \bibinfo {author} {\bibfnamefont {T.}~\bibnamefont {Ohashi}}, \ and\ \bibinfo
  {author} {\bibfnamefont {S.-i.}\ \bibnamefont {Suga}},\ }\href {\doibase
  10.1143/JPSJ.77.094707} {\bibfield  {journal} {\bibinfo  {journal} {Journal
  of the Physical Society of Japan}\ }\textbf {\bibinfo {volume} {77}},\
  \bibinfo {pages} {094707} (\bibinfo {year} {2008})}\BibitemShut {NoStop}%
\bibitem [{\citenamefont {Kuzmenko}\ and\ \citenamefont
  {Avishai}(2014)}]{su12}%
  \BibitemOpen
  \bibfield  {author} {\bibinfo {author} {\bibfnamefont {I.}~\bibnamefont
  {Kuzmenko}}\ and\ \bibinfo {author} {\bibfnamefont {Y.}~\bibnamefont
  {Avishai}},\ }\href {\doibase 10.1103/PhysRevB.89.195110} {\bibfield
  {journal} {\bibinfo  {journal} {Phys. Rev. B}\ }\textbf {\bibinfo {volume}
  {89}},\ \bibinfo {pages} {195110} (\bibinfo {year} {2014})}\BibitemShut
  {NoStop}%
\bibitem [{\citenamefont {Nishida}(2013)}]{nashida1}%
  \BibitemOpen
  \bibfield  {author} {\bibinfo {author} {\bibfnamefont {Y.}~\bibnamefont
  {Nishida}},\ }\href {\doibase 10.1103/PhysRevLett.111.135301} {\bibfield
  {journal} {\bibinfo  {journal} {Phys. Rev. Lett.}\ }\textbf {\bibinfo
  {volume} {111}},\ \bibinfo {pages} {135301} (\bibinfo {year}
  {2013})}\BibitemShut {NoStop}%
\bibitem [{\citenamefont {Bauer}\ \emph {et~al.}(2013)\citenamefont {Bauer},
  \citenamefont {Salomon},\ and\ \citenamefont {Demler}}]{salamon}%
  \BibitemOpen
  \bibfield  {author} {\bibinfo {author} {\bibfnamefont {J.}~\bibnamefont
  {Bauer}}, \bibinfo {author} {\bibfnamefont {C.}~\bibnamefont {Salomon}}, \
  and\ \bibinfo {author} {\bibfnamefont {E.}~\bibnamefont {Demler}},\ }\href
  {\doibase 10.1103/PhysRevLett.111.215304} {\bibfield  {journal} {\bibinfo
  {journal} {Phys. Rev. Lett.}\ }\textbf {\bibinfo {volume} {111}},\ \bibinfo
  {pages} {215304} (\bibinfo {year} {2013})}\BibitemShut {NoStop}%
\bibitem [{\citenamefont {Nishida}(2016)}]{nashida2}%
  \BibitemOpen
  \bibfield  {author} {\bibinfo {author} {\bibfnamefont {Y.}~\bibnamefont
  {Nishida}},\ }\href {\doibase 10.1103/PhysRevA.93.011606} {\bibfield
  {journal} {\bibinfo  {journal} {Phys. Rev. A}\ }\textbf {\bibinfo {volume}
  {93}},\ \bibinfo {pages} {011606} (\bibinfo {year} {2016})}\BibitemShut
  {NoStop}%
\bibitem [{\citenamefont {Kuzmenko}\ \emph {et~al.}(2016)\citenamefont
  {Kuzmenko}, \citenamefont {Kuzmenko}, \citenamefont {Avishai},\ and\
  \citenamefont {Jo}}]{su6}%
  \BibitemOpen
  \bibfield  {author} {\bibinfo {author} {\bibfnamefont {I.}~\bibnamefont
  {Kuzmenko}}, \bibinfo {author} {\bibfnamefont {T.}~\bibnamefont {Kuzmenko}},
  \bibinfo {author} {\bibfnamefont {Y.}~\bibnamefont {Avishai}}, \ and\
  \bibinfo {author} {\bibfnamefont {G.-B.}\ \bibnamefont {Jo}},\ }\href
  {\doibase 10.1103/PhysRevB.93.115143} {\bibfield  {journal} {\bibinfo
  {journal} {Phys. Rev. B}\ }\textbf {\bibinfo {volume} {93}},\ \bibinfo
  {pages} {115143} (\bibinfo {year} {2016})}\BibitemShut {NoStop}%
\bibitem [{\citenamefont {Mora}\ \emph {et~al.}(2009)\citenamefont {Mora},
  \citenamefont {Vitushinsky}, \citenamefont {Leyronas}, \citenamefont
  {Clerk},\ and\ \citenamefont {Le~Hur}}]{mora1}%
  \BibitemOpen
  \bibfield  {author} {\bibinfo {author} {\bibfnamefont {C.}~\bibnamefont
  {Mora}}, \bibinfo {author} {\bibfnamefont {P.}~\bibnamefont {Vitushinsky}},
  \bibinfo {author} {\bibfnamefont {X.}~\bibnamefont {Leyronas}}, \bibinfo
  {author} {\bibfnamefont {A.~A.}\ \bibnamefont {Clerk}}, \ and\ \bibinfo
  {author} {\bibfnamefont {K.}~\bibnamefont {Le~Hur}},\ }\href {\doibase
  10.1103/PhysRevB.80.155322} {\bibfield  {journal} {\bibinfo  {journal} {Phys.
  Rev. B}\ }\textbf {\bibinfo {volume} {80}},\ \bibinfo {pages} {155322}
  (\bibinfo {year} {2009})}\BibitemShut {NoStop}%
\bibitem [{\citenamefont {Delagrange}\ \emph {et~al.}(2018)\citenamefont
  {Delagrange}, \citenamefont {Basset}, \citenamefont {Bouchiat},\ and\
  \citenamefont {Deblock}}]{ll3}%
  \BibitemOpen
  \bibfield  {author} {\bibinfo {author} {\bibfnamefont {R.}~\bibnamefont
  {Delagrange}}, \bibinfo {author} {\bibfnamefont {J.}~\bibnamefont {Basset}},
  \bibinfo {author} {\bibfnamefont {H.}~\bibnamefont {Bouchiat}}, \ and\
  \bibinfo {author} {\bibfnamefont {R.}~\bibnamefont {Deblock}},\ }\href
  {\doibase 10.1103/PhysRevB.97.041412} {\bibfield  {journal} {\bibinfo
  {journal} {Phys. Rev. B}\ }\textbf {\bibinfo {volume} {97}},\ \bibinfo
  {pages} {041412} (\bibinfo {year} {2018})}\BibitemShut {NoStop}%
\bibitem [{\citenamefont {Anderson}(1961)}]{and1}%
  \BibitemOpen
  \bibfield  {author} {\bibinfo {author} {\bibfnamefont {P.~W.}\ \bibnamefont
  {Anderson}},\ }\href {\doibase 10.1103/PhysRev.124.41} {\bibfield  {journal}
  {\bibinfo  {journal} {Phys. Rev.}\ }\textbf {\bibinfo {volume} {124}},\
  \bibinfo {pages} {41} (\bibinfo {year} {1961})}\BibitemShut {NoStop}%
\bibitem [{\citenamefont {Krishna-murthy}\ \emph {et~al.}(1980)\citenamefont
  {Krishna-murthy}, \citenamefont {Wilkins},\ and\ \citenamefont
  {Wilson}}]{and2}%
  \BibitemOpen
  \bibfield  {author} {\bibinfo {author} {\bibfnamefont {H.~R.}\ \bibnamefont
  {Krishna-murthy}}, \bibinfo {author} {\bibfnamefont {J.~W.}\ \bibnamefont
  {Wilkins}}, \ and\ \bibinfo {author} {\bibfnamefont {K.~G.}\ \bibnamefont
  {Wilson}},\ }\href {\doibase 10.1103/PhysRevB.21.1003} {\bibfield  {journal}
  {\bibinfo  {journal} {Phys. Rev. B}\ }\textbf {\bibinfo {volume} {21}},\
  \bibinfo {pages} {1003} (\bibinfo {year} {1980})}\BibitemShut {NoStop}%
\bibitem [{\citenamefont {Glazman}\ and\ \citenamefont {Raikh}(1988)}]{GR}%
  \BibitemOpen
  \bibfield  {author} {\bibinfo {author} {\bibfnamefont {L.~I.}\ \bibnamefont
  {Glazman}}\ and\ \bibinfo {author} {\bibfnamefont {M.~E.}\ \bibnamefont
  {Raikh}},\ }\href@noop {} {\bibfield  {journal} {\bibinfo  {journal} {J. Exp.
  Theor. Phys.}\ }\textbf {\bibinfo {volume} {27}},\ \bibinfo {pages} {452}
  (\bibinfo {year} {1988})}\BibitemShut {NoStop}%
\bibitem [{\citenamefont {Schrieffer}\ and\ \citenamefont {Wolff}(1966)}]{sw}%
  \BibitemOpen
  \bibfield  {author} {\bibinfo {author} {\bibfnamefont {J.~R.}\ \bibnamefont
  {Schrieffer}}\ and\ \bibinfo {author} {\bibfnamefont {P.~A.}\ \bibnamefont
  {Wolff}},\ }\href {\doibase 10.1103/PhysRev.149.491} {\bibfield  {journal}
  {\bibinfo  {journal} {Phys. Rev.}\ }\textbf {\bibinfo {volume} {149}},\
  \bibinfo {pages} {491} (\bibinfo {year} {1966})}\BibitemShut {NoStop}%
\bibitem [{\citenamefont {Parcollet}\ \emph {et~al.}(1998)\citenamefont
  {Parcollet}, \citenamefont {Georges}, \citenamefont {Kotliar},\ and\
  \citenamefont {Sengupta}}]{t}%
  \BibitemOpen
  \bibfield  {author} {\bibinfo {author} {\bibfnamefont {O.}~\bibnamefont
  {Parcollet}}, \bibinfo {author} {\bibfnamefont {A.}~\bibnamefont {Georges}},
  \bibinfo {author} {\bibfnamefont {G.}~\bibnamefont {Kotliar}}, \ and\
  \bibinfo {author} {\bibfnamefont {A.}~\bibnamefont {Sengupta}},\ }\href
  {\doibase 10.1103/PhysRevB.58.3794} {\bibfield  {journal} {\bibinfo
  {journal} {Phys. Rev. B}\ }\textbf {\bibinfo {volume} {58}},\ \bibinfo
  {pages} {3794} (\bibinfo {year} {1998})}\BibitemShut {NoStop}%
\bibitem [{\citenamefont {Mora}(2009)}]{mora2}%
  \BibitemOpen
  \bibfield  {author} {\bibinfo {author} {\bibfnamefont {C.}~\bibnamefont
  {Mora}},\ }\href
  {https://journals.aps.org/prb/abstract/10.1103/PhysRevB.80.125304} {\bibfield
   {journal} {\bibinfo  {journal} {Phys. Rev. B}\ }\textbf {\bibinfo {volume}
  {80}},\ \bibinfo {pages} {125304} (\bibinfo {year} {2009})}\BibitemShut
  {NoStop}%
\bibitem [{\citenamefont {Nozi{\'e}res}(1974)}]{Nozieres}%
  \BibitemOpen
  \bibfield  {author} {\bibinfo {author} {\bibfnamefont {P.}~\bibnamefont
  {Nozi{\'e}res}},\ }\href {https://doi.org/10.1007/BF00654541} {\bibfield
  {journal} {\bibinfo  {journal} {J. Low Temp. Phys.}\ }\textbf {\bibinfo
  {volume} {17}},\ \bibinfo {pages} {31} (\bibinfo {year} {1974})}\BibitemShut
  {NoStop}%
\bibitem [{\citenamefont {Affleck}\ and\ \citenamefont {Ludwig}(1993)}]{aff}%
  \BibitemOpen
  \bibfield  {author} {\bibinfo {author} {\bibfnamefont {I.}~\bibnamefont
  {Affleck}}\ and\ \bibinfo {author} {\bibfnamefont {A.~W.~W.}\ \bibnamefont
  {Ludwig}},\ }\href
  {https://journals.aps.org/prb/abstract/10.1103/PhysRevB.48.7297} {\bibfield
  {journal} {\bibinfo  {journal} {Phys. Rev. B}\ }\textbf {\bibinfo {volume}
  {48}},\ \bibinfo {pages} {7297} (\bibinfo {year} {1993})}\BibitemShut
  {NoStop}%
\bibitem [{\citenamefont {Cox}\ and\ \citenamefont {Zawadowski}(1998)}]{Cox}%
  \BibitemOpen
  \bibfield  {author} {\bibinfo {author} {\bibfnamefont {D.~L.}\ \bibnamefont
  {Cox}}\ and\ \bibinfo {author} {\bibfnamefont {A.}~\bibnamefont
  {Zawadowski}},\ }\href {https://doi.org/10.1080/000187398243500} {\bibfield
  {journal} {\bibinfo  {journal} {Advances in Physics}\ }\textbf {\bibinfo
  {volume} {47}},\ \bibinfo {pages} {599} (\bibinfo {year} {1998})}\BibitemShut
  {NoStop}%
\bibitem [{\citenamefont {Karki}\ \emph {et~al.}(2018)\citenamefont {Karki},
  \citenamefont {Mora}, \citenamefont {von Delft},\ and\ \citenamefont
  {Kiselev}}]{dee2}%
  \BibitemOpen
  \bibfield  {author} {\bibinfo {author} {\bibfnamefont {D.~B.}\ \bibnamefont
  {Karki}}, \bibinfo {author} {\bibfnamefont {C.}~\bibnamefont {Mora}},
  \bibinfo {author} {\bibfnamefont {J.}~\bibnamefont {von Delft}}, \ and\
  \bibinfo {author} {\bibfnamefont {M.~N.}\ \bibnamefont {Kiselev}},\ }\href
  {\doibase 10.1103/PhysRevB.97.195403} {\bibfield  {journal} {\bibinfo
  {journal} {Phys. Rev. B}\ }\textbf {\bibinfo {volume} {97}},\ \bibinfo
  {pages} {195403} (\bibinfo {year} {2018})}\BibitemShut {NoStop}%
\bibitem [{\citenamefont {Karki}\ and\ \citenamefont {Kiselev}(2018)}]{dee3}%
  \BibitemOpen
  \bibfield  {author} {\bibinfo {author} {\bibfnamefont {D.~B.}\ \bibnamefont
  {Karki}}\ and\ \bibinfo {author} {\bibfnamefont {M.~N.}\ \bibnamefont
  {Kiselev}},\ }\href {\doibase 10.1103/PhysRevB.98.165443} {\bibfield
  {journal} {\bibinfo  {journal} {Phys. Rev. B}\ }\textbf {\bibinfo {volume}
  {98}},\ \bibinfo {pages} {165443} (\bibinfo {year} {2018})}\BibitemShut
  {NoStop}%
\bibitem [{\citenamefont {Keldysh}(1965)}]{keldysh}%
  \BibitemOpen
  \bibfield  {author} {\bibinfo {author} {\bibfnamefont {L.~V.}\ \bibnamefont
  {Keldysh}},\ }\href
  {http://www.jetp.ac.ru/cgi-bin/e/index/e/20/4/p1018?a=list} {\bibfield
  {journal} {\bibinfo  {journal} {Sov. Phys. JETP}\ }\textbf {\bibinfo {volume}
  {20}},\ \bibinfo {pages} {1018} (\bibinfo {year} {1965})}\BibitemShut
  {NoStop}%
\bibitem [{\citenamefont {Kim}\ and\ \citenamefont {Hershfield}(2002)}]{kimh}%
  \BibitemOpen
  \bibfield  {author} {\bibinfo {author} {\bibfnamefont {T.-S.}\ \bibnamefont
  {Kim}}\ and\ \bibinfo {author} {\bibfnamefont {S.}~\bibnamefont
  {Hershfield}},\ }\href {\doibase 10.1103/PhysRevLett.88.136601} {\bibfield
  {journal} {\bibinfo  {journal} {Phys. Rev. Lett.}\ }\textbf {\bibinfo
  {volume} {88}},\ \bibinfo {pages} {136601} (\bibinfo {year}
  {2002})}\BibitemShut {NoStop}%
\bibitem [{\citenamefont {Kim}\ and\ \citenamefont {Hershfield}(2003)}]{kim}%
  \BibitemOpen
  \bibfield  {author} {\bibinfo {author} {\bibfnamefont {T.-S.}\ \bibnamefont
  {Kim}}\ and\ \bibinfo {author} {\bibfnamefont {S.}~\bibnamefont
  {Hershfield}},\ }\href {\doibase 10.1103/PhysRevB.67.165313} {\bibfield
  {journal} {\bibinfo  {journal} {Phys. Rev. B}\ }\textbf {\bibinfo {volume}
  {67}},\ \bibinfo {pages} {165313} (\bibinfo {year} {2003})}\BibitemShut
  {NoStop}%
\bibitem [{\citenamefont {Beenakker}\ and\ \citenamefont
  {Staring}(1992)}]{ben}%
  \BibitemOpen
  \bibfield  {author} {\bibinfo {author} {\bibfnamefont {C.~W.~J.}\
  \bibnamefont {Beenakker}}\ and\ \bibinfo {author} {\bibfnamefont {A.~A.~M.}\
  \bibnamefont {Staring}},\ }\href {\doibase 10.1103/PhysRevB.46.9667}
  {\bibfield  {journal} {\bibinfo  {journal} {Phys. Rev. B}\ }\textbf {\bibinfo
  {volume} {46}},\ \bibinfo {pages} {9667} (\bibinfo {year}
  {1992})}\BibitemShut {NoStop}%
\bibitem [{\citenamefont {Zlatic}\ and\ \citenamefont
  {Monnier}(2014)}]{Zlatic}%
  \BibitemOpen
  \bibfield  {author} {\bibinfo {author} {\bibfnamefont {V.}~\bibnamefont
  {Zlatic}}\ and\ \bibinfo {author} {\bibfnamefont {R.}~\bibnamefont
  {Monnier}},\ }\href@noop {} {\emph {\bibinfo {title} {Modern Theory of
  Thermoelectricity}}}\ (\bibinfo  {publisher} {Oxford University Press},\
  \bibinfo {year} {2014})\BibitemShut {NoStop}%
\bibitem [{\citenamefont {Matveev}\ and\ \citenamefont {Andreev}(2002)}]{mat}%
  \BibitemOpen
  \bibfield  {author} {\bibinfo {author} {\bibfnamefont {K.~A.}\ \bibnamefont
  {Matveev}}\ and\ \bibinfo {author} {\bibfnamefont {A.~V.}\ \bibnamefont
  {Andreev}},\ }\href {\doibase 10.1103/PhysRevB.66.045301} {\bibfield
  {journal} {\bibinfo  {journal} {Phys. Rev. B}\ }\textbf {\bibinfo {volume}
  {66}},\ \bibinfo {pages} {045301} (\bibinfo {year} {2002})}\BibitemShut
  {NoStop}%
\bibitem [{\citenamefont {Carmi}\ \emph {et~al.}(2012)\citenamefont {Carmi},
  \citenamefont {Oreg}, \citenamefont {Berkooz},\ and\ \citenamefont
  {Goldhaber-Gordon}}]{carmi1}%
  \BibitemOpen
  \bibfield  {author} {\bibinfo {author} {\bibfnamefont {A.}~\bibnamefont
  {Carmi}}, \bibinfo {author} {\bibfnamefont {Y.}~\bibnamefont {Oreg}},
  \bibinfo {author} {\bibfnamefont {M.}~\bibnamefont {Berkooz}}, \ and\
  \bibinfo {author} {\bibfnamefont {D.}~\bibnamefont {Goldhaber-Gordon}},\
  }\href {\doibase 10.1103/PhysRevB.86.115129} {\bibfield  {journal} {\bibinfo
  {journal} {Phys. Rev. B}\ }\textbf {\bibinfo {volume} {86}},\ \bibinfo
  {pages} {115129} (\bibinfo {year} {2012})}\BibitemShut {NoStop}%
\bibitem [{\citenamefont {Pustilnik}\ and\ \citenamefont
  {Glazman}(2001)}]{lea}%
  \BibitemOpen
  \bibfield  {author} {\bibinfo {author} {\bibfnamefont {M.}~\bibnamefont
  {Pustilnik}}\ and\ \bibinfo {author} {\bibfnamefont {L.~I.}\ \bibnamefont
  {Glazman}},\ }\href {\doibase 10.1103/PhysRevLett.87.216601} {\bibfield
  {journal} {\bibinfo  {journal} {Phys. Rev. Lett.}\ }\textbf {\bibinfo
  {volume} {87}},\ \bibinfo {pages} {216601} (\bibinfo {year}
  {2001})}\BibitemShut {NoStop}%
\bibitem [{\citenamefont {Pustilnik}\ and\ \citenamefont
  {Glazman}(2004)}]{Pust}%
  \BibitemOpen
  \bibfield  {author} {\bibinfo {author} {\bibfnamefont {M.}~\bibnamefont
  {Pustilnik}}\ and\ \bibinfo {author} {\bibfnamefont {L.}~\bibnamefont
  {Glazman}},\ }\href {\doibase 10.1088/0953-8984/16/16/r01} {\bibfield
  {journal} {\bibinfo  {journal} {Journal of Physics: Condensed Matter}\
  }\textbf {\bibinfo {volume} {16}},\ \bibinfo {pages} {R513} (\bibinfo {year}
  {2004})}\BibitemShut {NoStop}%
\bibitem [{\citenamefont {Pustilnik}\ \emph {et~al.}(2004)\citenamefont
  {Pustilnik}, \citenamefont {Borda}, \citenamefont {Glazman},\ and\
  \citenamefont {von Delft}}]{jvd}%
  \BibitemOpen
  \bibfield  {author} {\bibinfo {author} {\bibfnamefont {M.}~\bibnamefont
  {Pustilnik}}, \bibinfo {author} {\bibfnamefont {L.}~\bibnamefont {Borda}},
  \bibinfo {author} {\bibfnamefont {L.~I.}\ \bibnamefont {Glazman}}, \ and\
  \bibinfo {author} {\bibfnamefont {J.}~\bibnamefont {von Delft}},\ }\href
  {\doibase 10.1103/PhysRevB.69.115316} {\bibfield  {journal} {\bibinfo
  {journal} {Phys. Rev. B}\ }\textbf {\bibinfo {volume} {69}},\ \bibinfo
  {pages} {115316} (\bibinfo {year} {2004})}\BibitemShut {NoStop}%
\bibitem [{\citenamefont {Park}\ \emph {et~al.}(2013)\citenamefont {Park},
  \citenamefont {Lee}, \citenamefont {Oreg},\ and\ \citenamefont {Sim}}]{yg}%
  \BibitemOpen
  \bibfield  {author} {\bibinfo {author} {\bibfnamefont {J.}~\bibnamefont
  {Park}}, \bibinfo {author} {\bibfnamefont {S.-S.~B.}\ \bibnamefont {Lee}},
  \bibinfo {author} {\bibfnamefont {Y.}~\bibnamefont {Oreg}}, \ and\ \bibinfo
  {author} {\bibfnamefont {H.-S.}\ \bibnamefont {Sim}},\ }\href {\doibase
  10.1103/PhysRevLett.110.246603} {\bibfield  {journal} {\bibinfo  {journal}
  {Phys. Rev. Lett.}\ }\textbf {\bibinfo {volume} {110}},\ \bibinfo {pages}
  {246603} (\bibinfo {year} {2013})}\BibitemShut {NoStop}%
\bibitem [{\citenamefont {Dorda}\ \emph {et~al.}(2016)\citenamefont {Dorda},
  \citenamefont {Ganahl}, \citenamefont {Andergassen}, \citenamefont {von~der
  Linden},\ and\ \citenamefont {Arrigoni}}]{sabina}%
  \BibitemOpen
  \bibfield  {author} {\bibinfo {author} {\bibfnamefont {A.}~\bibnamefont
  {Dorda}}, \bibinfo {author} {\bibfnamefont {M.}~\bibnamefont {Ganahl}},
  \bibinfo {author} {\bibfnamefont {S.}~\bibnamefont {Andergassen}}, \bibinfo
  {author} {\bibfnamefont {W.}~\bibnamefont {von~der Linden}}, \ and\ \bibinfo
  {author} {\bibfnamefont {E.}~\bibnamefont {Arrigoni}},\ }\href {\doibase
  10.1103/PhysRevB.94.245125} {\bibfield  {journal} {\bibinfo  {journal} {Phys.
  Rev. B}\ }\textbf {\bibinfo {volume} {94}},\ \bibinfo {pages} {245125}
  (\bibinfo {year} {2016})}\BibitemShut {NoStop}%
\bibitem [{\citenamefont {P\'erez~Daroca}\ \emph {et~al.}(2018)\citenamefont
  {P\'erez~Daroca}, \citenamefont {Roura-Bas},\ and\ \citenamefont
  {Aligia}}]{aligia}%
  \BibitemOpen
  \bibfield  {author} {\bibinfo {author} {\bibfnamefont {D.}~\bibnamefont
  {P\'erez~Daroca}}, \bibinfo {author} {\bibfnamefont {P.}~\bibnamefont
  {Roura-Bas}}, \ and\ \bibinfo {author} {\bibfnamefont {A.~A.}\ \bibnamefont
  {Aligia}},\ }\href {\doibase 10.1103/PhysRevB.97.165433} {\bibfield
  {journal} {\bibinfo  {journal} {Phys. Rev. B}\ }\textbf {\bibinfo {volume}
  {97}},\ \bibinfo {pages} {165433} (\bibinfo {year} {2018})}\BibitemShut
  {NoStop}%
\bibitem [{\citenamefont {{Eckern}}\ and\ \citenamefont
  {{Wysoki{\'n}ski}}(2019)}]{ue}%
  \BibitemOpen
  \bibfield  {author} {\bibinfo {author} {\bibfnamefont {U.}~\bibnamefont
  {{Eckern}}}\ and\ \bibinfo {author} {\bibfnamefont {K.~I.}\ \bibnamefont
  {{Wysoki{\'n}ski}}},\ }\href@noop {} {\bibfield  {journal} {\bibinfo
  {journal} {arXiv e-prints}\ ,\ \bibinfo {eid} {arXiv:1904.05064}} (\bibinfo
  {year} {2019})}\BibitemShut {NoStop}%
\end{thebibliography}
\end{document}